\documentclass{iopart}
\usepackage{graphicx}  
\usepackage{latexsym}  

\jl{1}        
\eqnobysec    

\def\beq{\begin{equation}}
\def\eeq{\end{equation}}
\def\rmd{{\rm d}}

\begin{document}

\title[Spinning test particles in Weyl spacetimes]
{Spinning test particles in Weyl spacetimes}

\author{
Donato Bini$^{*\|\P}$, 
Fernando de Felice$^\dagger$, 
Andrea Geralico$^{\ddag\|}$ 
and 
Andrea Lunari$^{\S\|}$}
\address{
  ${}^*$\
Istituto per le Applicazioni del Calcolo ``M. Picone'', CNR I-00161 Rome, Italy
}
\address{
  ${}^\|$\
  International Center for Relativistic Astrophysics,
  University of Rome, I-00185 Rome, Italy
}
\address{
${}^\P$
  INFN - Sezione di Firenze, Polo Scientifico, Via Sansone 1, 
  I-50019, Sesto Fiorentino (FI), Italy 
}
\address{
${}^\dagger$\
Dipartimento di Fisica, Universit\`a di Padova, and INFN, Sezione di Padova, Via Marzolo 8,  I-35131 Padova, Italy
}
\address{
  ${}^\ddag$\
  Dipartimento di Fisica, Universit\`a di Lecce, and INFN - Sezione di Lecce,
  Via Arnesano, CP 193, I-73100 Lecce, Italy
}
\address{
  ${}^\S$\
  Dipartimento di Fisica, Universit\`a dell'Insubria, I-22100 Como, Italy, and INFN - Sezione di Milano
}

\begin{abstract}
The motion of spinning test particles along circular orbits in static vacuum spacetimes belonging to the Weyl class is discussed.
Spin alignment and coupling with background parameters in the case of superimposed Weyl fields, corresponding to a single Schwarzschild black hole and single Chazy-Curzon particle as well as to two Schwarzschild black holes and two Chazy-Curzon particles, are studied in detail for standard choices of supplementary conditions.
Applications to the gravitomagnetic \lq\lq clock effect'' are also discussed.
\end{abstract}

\pacno{04.20.Cv}

\submitto{\JPA}

\section{Introduction}

The study of spinning test particles in General Relativity started long ago after the pioneering works of Mathisson and Papapetrou \cite{math37,papa51}. 
The standard model for the description of spinning test particles in General Relativity is actually known as the Mathisson-Papapetrou model and it consists in a set of 10 partial differential equations for 13 unknown variables
needed to describe the spinning test particle, i.e. the (timelike) generalized 4-momentum of the particle $P$, the (antisymmetric) spin 2-tensor $S$ and the unit timelike vector $U$ tangent to the world line used to perform a multipole moments reduction, truncated to the first order to define a spin structure for the particle.
To complete the scheme three further conditions relating $P$, $S$ and $U$ are necessary and there exist natural choices for this, widely discussed in the literature \cite{cori51,pir56,tulc59}. Detailed  studies concerning spinning test particles in General Relativity are due to Dixon \cite{dixon64,dixon69,dixon70,dixon73,dixon74}, 
Taub \cite{taub64}, Mashhoon \cite{masspin1,masspin2} and Ehlers and Rudolph \cite{ehlers77}.

Recent claims concerning the various couplings between spin and rotation \cite{mash88},
spin and acceleration \cite{bcm} have started new interest and motivated further investigations from both theoretical and experimental point of view.

In this paper we study the motion of spinning test particles on circular orbits in static vacuum spacetimes belonging to the Weyl class \cite{weyl,exactsols}, with explicit example for one-body and two-body solutions consisting of Schwarzschild black holes and Chazy-Curzon particles, and compare the results obtained by using different supplementary conditions, generalizing the pioneering work of Tod, de Felice and Calvani \cite{tod77}, and 
the most recent ones by Bini, de Felice and Geralico \cite{bdfg1,bdfg2}.

The paper is organized as follows. In Section 2 we review the properties of timelike spatially circular orbits followed by non-spinning test particles in axisymmetric static vacuum spacetimes \cite{bgijmpd}. The Mathisson-Papapetrou equations of motion for spinning test particles in circular motion are introduced in Section 3, assuming constant frame components for the spin tensor with respect to a frame adapted to the symmetries of the spacetime. The solution is then characterized in the subsequent subsections in terms of the various possible choices of Corinaldesi and Papapetrou, Pirani and Tulczyjew supplementary conditions; the limiting situation of small spin is described too. In Section 4 the so called gravitomagnetic \lq\lq clock effect" due to the difference in the arrival times of two oppositely rotating orbits is deduced when the motion is confined on particular symmetry hyperplanes. 
Finally, applications to specific Weyl spacetimes are discussed in Section 5, in order to make more concrete the whole treatment. 
 
In what follows Greek indices run from 0 to 3 while Latin indeces run from 1 to 3; the spacetime metric signature is +2 and geometrized units are used such that both the velocity of light in  vacuum $c$ and the gravitational constant $G$ are set equal to one.

\section{Vacuum Weyl spacetimes and circular orbits}

Spatially circular orbits are most important in astrophysics; therefore we judge it useful to make available all the geometrical properties of those orbits in the Weyl class of spacetimes. As well known, axisymmetric, static, vacuum solutions of the Einstein's field equations can be described by the Weyl formalism \cite{weyl}. The corresponding line element in Weyl  canonical coordinates \cite{exactsols} $(x^0=t,x^1=\rho,x^2=z,x^3=\phi)$ is 
\begin{equation}
\label{weylmetric}
\rmd s^2=-e^{2\psi}\rmd t^2+e^{2(\gamma-\psi)}[\rmd \rho^2+\rmd z^2]+\rho^2e^{-2\psi}\rmd \phi^2\ ,
\end{equation}
where the function $\psi$ and $\gamma$ depend on coordinates $\rho$ and $z$ only. The vacuum Einstein's field equations reduce to the following: 
\begin{eqnarray}
\label{einsteqs}
&& \qquad \qquad \psi_{,\rho\rho}+\frac1\rho\psi_{,\rho}+\psi_{,zz}=0, \nonumber\\
&& \gamma_{,\rho}-\rho[\psi_{,\rho}^2-\psi_{,z}^2]=0, \qquad
\gamma_{,z}-2\rho\psi_{,\rho}\psi_{,z}=0\ .
\end{eqnarray}
It is useful to introduce the orthonormal frame 
\begin{equation}
e_{\hat t}=e^{-\psi}\partial_t\ , \quad e_{\hat \rho}=e^{\psi-\gamma}\partial_\rho\ , \quad e_{\hat z}=e^{\psi-\gamma}\partial_z\ , \quad e_{\hat \phi}=\frac{e^\psi}{\rho}\partial_\phi\ ,
\end{equation}
with dual frame
\begin{equation}
\label{of_zamo}
\omega^{{\hat t}}=e^{\psi}\rmd t\ , \quad \omega^{{\hat \rho}}=e^{\gamma-\psi}\rmd \rho\ , \quad 
\omega^{{\hat z}}=e^{\gamma-\psi}\rmd z\ , \quad \omega^{{\hat \phi}}=\rho e^{-\psi}\rmd \phi\ .
\end{equation}
In the metric (\ref{weylmetric}) let us consider a family of test particles spatially moving along the $\phi$ direction with constant speed; the (timelike) 4-velocity $U$ associated to a generic orbit within the family is the following:
\begin{equation}
\label{circolare}
U=\Gamma_{\zeta}[\partial_t + \zeta\partial_{\phi}]=\gamma_n [n +\nu  e_{\hat\phi}]=\cosh \alpha\,  n +\sinh \alpha\,  e_{\hat\phi}\ ,
\end{equation}
where $n\equiv e_{\hat t}$ denotes the 4-velocity of the standard family of static observers; $\gamma_n=-U\cdot n=(1-\nu^2)^{-1/2}=\Gamma_\zeta e^\psi$ is the Lorentz factor and $\zeta$, $\nu$ or $\alpha$ are respectively the angular velocity, the speed and the rapidity parametrization of the whole family.
They are all constant along the orbit and satisfy the mutual relations
\begin{equation}
\nu=e^{-2\psi}\,\rho \zeta=\tanh \alpha\ .
\end{equation}
$\Gamma_{\zeta}$ is defined by the timelike condition $U\cdot U=-1$ as
\begin{equation}
-\Gamma_{\zeta}^{-2}=g_{tt}+\zeta^2g_{\phi\phi}=-e^{2\psi}+\zeta^2\rho^2e^{-2\psi}=-\frac{e^{2\psi}}{\gamma_n^2}\ .
\end{equation}
Moreover, the physical dimension of $\zeta$ is $[\rm lenght]^{-1}$, while $\nu$, $\gamma_n$ and $\Gamma_{\zeta}$ are pure 
numbers \footnote{The relative dimensions of the 4-vector $U$ vary with the
components consistently with the dimensions of the corresponding
coordinate basis vector.
Hence $[U^t]=1$ and $[U^\phi]=[\rm length]^{-1}$.
In an analogous way, the 4-vector $n=e^{-\psi}\partial_t$ has for its
only component the  dimensions $[n^t]=1$.
From the above it follows correctly that $[\gamma_n]=[e^\psi U^t]=1.$ }.

For late purposes, it is useful to introduce the (spacelike) unit vector $E_{\hat \phi}$ orthogonal to $U$ in the $(t, \phi)$ plane, obtained by boosting 
$e_{\hat \phi}$ in the local rest space of $U$
\begin{equation}\fl\quad
\label{barU}
E_{\hat \phi}={\bar \Gamma}_{\bar \zeta}[\partial_t + {\bar \zeta}\partial_{\phi}]=\gamma_n (\nu n +e_{\hat \phi}) , \qquad {\bar \Gamma}_{\bar \zeta}=\Gamma_{\zeta}\left[\frac{\zeta}{{\bar \zeta}}\right]^{1/2}\ , \qquad {\bar \zeta}
=-\frac1{\zeta}\frac{e^{4\psi}}{\rho^2}\ ,
\end{equation}
so that $E_{\hat \phi}\cdot E_{\hat \phi}=1$ and $E_{\hat \phi}\cdot U=0$.

It is worth noting also that the case of null orbits for rotating photons corresponds to 
\begin{equation}
\zeta_{{\rm null}\ \pm}=\pm \frac{e^{2\psi}}{\rho}, \qquad \nu_{{\rm null}\ \pm}=\pm 1.
\end{equation}

The non-vanishing components of the 4-acceleration $a(U)=\nabla_U U$ of $U$ are given by
\begin{eqnarray}
\label{acccomp}\fl\quad
a(U)^{\hat \rho}&=&\frac{e^{\psi-\gamma}}{e^{4\psi}-\rho^2\zeta^2}\left[
\psi_{,\rho} (e^{4\psi}+\rho^2\zeta^2)-\rho\zeta^2\right]=e^{\psi-\gamma} \gamma_n^2 [\psi_{,\rho} -\frac{\nu^2}{\rho}(1-\rho \psi_{,\rho})] ,\nonumber \\
\fl\quad
a(U)^{\hat z}&=&e^{\psi-\gamma}\psi_{,z}\frac{e^{4\psi}+\rho^2\zeta^2}{e^{4\psi}-\rho^2\zeta^2}= e^{\psi-\gamma}\psi_{,z}\gamma_n^2 (1+\nu^2) \ .
\end{eqnarray}
They all have dimensions of lenght${}^{-1}$.
The absolute value of the acceleration is
\begin{equation}\fl\quad
\label{modacc}
\kappa=||a(U)||=e^{\psi-\gamma} \gamma_n^2 \left[ \left(\psi_{,\rho} -\frac{\nu^2}{\rho}(1-\rho \psi_{,\rho})\right)^2
+\psi_{,z}^2 (1+\nu^2)^2 \right]^{1/2}\ ,
\end{equation}
and it is symmetric as a function of $\nu$;
one can also introduce polar coordinates in the acceleration plane $(\kappa, \chi)$
\begin{equation}\fl\quad
a(U)^{\hat \rho}=\kappa \cos\chi\ , \qquad 
a(U)^{\hat z}=\kappa \sin \chi\ , \qquad 
\tan \chi = \frac{\psi_{,z}(1+\nu^2)}{\psi_{,\rho} -\frac{\nu^2}{\rho}(1-\rho \psi_{,\rho})}\ ,
\end{equation}
so that the unit vector aligned with the acceleration is
\beq
\label{e1}
e_1=\cos \chi e_{\hat \rho}+\sin \chi e_{\hat z}\ ,
\eeq
and $\chi$ (depending on $\nu$ or $\alpha$) is constant along $U$: $\rmd \chi/\rmd \tau_U=0$.
Moreover both the unit vector associated with the spatial 3-velocity of the particle $U$ with respect to static observer $n$
\beq
\hat \nu(U,n)\equiv e_{\hat \phi}\ ,
\eeq
and (minus) the unit vector associated with the spatial 3-velocity of the observers $n$ with respect to particle $U$
\beq
[-\hat \nu(n,U)]\equiv E_{\hat \phi}
\eeq
can be used to define the relative curvatures \cite{bjdf0,bjdf,idcf1,idcf2} of the orbit of particle and observer.
This can be done by evaluating the derivatives along $U$ of the relative velocities unit vectors 
\beq
\frac{D e_{\hat \phi}}{\rmd \tau_U}\ , \qquad \frac{D E_{\hat \phi}}{\rmd \tau_U}\ , 
\eeq
which, in the spacetime metric under consideration (only), both belong to the $(\rho, z)$ plane, and define in turn
the centripetal acceleration 
\beq
\label{centrip}
\frac{D e_{\hat \phi}}{\rmd \tau_U}=a^{\rm (Centrip)}
\eeq
and the centrifugal one
\beq\label{centrif}
\frac{D E_{\hat \phi}}{\rmd \tau_U}=-a^{\rm (Centrif)}\ .
\eeq
For a complete discussion about centripetal and centrifugal forces in General Relativity see \cite{bjdf0,bjdf,idcf1,idcf2}, 
where the original definition uses a rescaling  of (\ref{centrip}) and (\ref{centrif}) by convenient factors of $\gamma_n$ and $\nu$.
Here we do not want to enter the discussion of what is the more appropriate definition  of centripetal and centrifugal forces; therefore,
these factors will not be introduced.
The result is 
$$a^{\rm (Centrip)}=\gamma_n \nu k_{\rm (lie)}\ ,$$ 
where $k_{\rm (lie)}= -\nabla\ln{(\sqrt{g_{\phi\phi}})}$, with components
\begin{equation}
\label{liecurv}
k_{\rm (lie)}{}_{\hat \rho} = - e^{\psi-\gamma}\frac{1-\rho \psi_{,\rho}}{\rho}\ , 
\qquad  k_{\rm (lie)}{}_{\hat z} = e^{\psi-\gamma}\psi_{,z}\ ,
\end{equation}
or, by using a polar representation for $k_{\rm (lie)}$, 
\begin{equation}\fl\quad
\label{kliepol}
k_{\rm (lie)}{}_{\hat \rho}=\kappa_{\rm (lie)}\cos \chi_{\rm (lie)}, \quad k_{\rm (lie)}{}_{\hat z}=\kappa_{\rm (lie)}\sin \chi_{\rm (lie)},\quad
\tan \chi_{\rm (lie)}= \frac{\rho \psi_{,z}}{\rho \psi_{,\rho}-1}\ .
\end{equation}
Moreover \footnote{The relative dimensions of the 4-vector $E_{\hat\phi}$ vary
according to its components. Specifically we have from its definition
$[E^t_{\hat\phi}]=1$ and $[E^\phi_{\hat\phi}]=[\rm lenght]^{-1}$. However the
dimensions of the 4-vector $DE_{\hat\phi}/d\tau_U$ depend on the
combination of the components of $U$, $E_{\hat\phi}$ and the connection
coefficients which appear in the covariant derivative; it turns out that
the only non zero components are
$$\left(\frac{DE_{\hat\phi}}{d\tau_U}\right)^{\hat \rho}\ , \qquad
\left(\frac{DE_{\hat\phi}}{d\tau_U}\right)^{\hat z} $$
and both have dimensions of $[\rm lenght]^{-2}$.
} 
\begin{equation}\label{DUdt}
-a^{\rm (Centrif)}=\frac{D E_{\hat \phi}}{\rmd \tau_U}=-e^{\psi-\gamma}\gamma_n^2\nu\left[\frac{2\rho\psi_{,\rho}-1}{\rho}e_{\hat \rho}+2\psi_{,z}e_{\hat z}\right]\ .
\end{equation}

The discussion presented above is quite standard now and it follows the notation of \cite{bjdf}. 
Special orbits can be selected so that
\begin{equation}
\label{nupm}
\zeta_{\pm}=\pm e^{2\psi}\left[\frac{\psi_{,\rho}}{\rho(1-\rho\psi_{,\rho})}\right]^{1/2}, \qquad
\nu_{\pm}=\pm \left[-1+\frac{1}{\rho\psi_{,\rho}}\right]^{-1/2}, 
\end{equation}
which permit the component $a(U)^{\hat \rho}$ to vanish; the quantities $\Gamma_{\zeta}$ and $a(U)^{\hat z}$ then become\footnote{The further requirement $a(U)^{\hat z}=0$ (and so $\psi_{,z}=0$) gives the conditions for circular geodesics.
These geodesics become null at the radius such that $\psi_{,\rho}=1/2\rho$.}
\begin{eqnarray}
\fl\quad
\Gamma_{\zeta_{\pm}}=e^{-\psi}\left[\frac{1-\rho\psi_{,\rho}}{1-2\rho\psi_{,\rho}}\right]^{1/2}\ , \qquad
a(U)^{\hat z}\big\vert_{\zeta=\zeta_{\pm}}=e^{\psi-\gamma}\frac{\psi_{,z}}{1-2\rho\psi_{,\rho}} \ .
\end{eqnarray}

Finally, by using $\nu_\pm$ and $k_{\rm (lie)}$ the components of the acceleration can be cast in the form
\begin{equation}
a(U)_{\hat \rho}=k_{\rm (lie)}{}_{\hat \rho}\, \gamma_n^2 (\nu^2-\nu_\pm^2), \qquad 
a(U)_{\hat z}= k_{\rm (lie)}{}_{\hat z} \, \gamma_n^2 (1+\nu^2)\ , 
\end{equation}
with the magnitude $\kappa$ (see equation~(\ref{modacc})) given by
\begin{equation}
\label{kappa}
\kappa=\gamma_n^2 \left[k_{\rm (lie)}{}_{\hat \rho}^2 (\nu^2-\nu_\pm^2)^2 + k_{\rm (lie)}{}_{\hat z}^2(1+\nu^2)^2\right]^{1/2}\ , 
\end{equation}
and the relation between $\chi$ and $\chi_{\rm (lie)}$ which becomes
\begin{equation}
\label{chilie}
\tan \chi = \frac{1+\nu^2}{\nu^2-\nu^2_\pm}\, \tan \chi_{\rm (lie)} . 
\end{equation}

Along each circular orbit one can set a Frenet-Serret (FS) frame \cite{iyer-vish} with $e_0=U$ and $e_1,e_2,e_3$ satisfying the system of evolution equations
\begin{eqnarray}
\label{FSeqs}
\frac{De_0}{\rmd\tau_U}&=&\kappa e_1,\qquad  \frac{De_1}{\rmd \tau_U}=\kappa e_0+\tau_1 e_2,\nonumber \\
 \nonumber \\
\frac{De_2}{\rmd \tau_U}&=&-\tau_1e_1+\tau_2e_3, \qquad \frac{De_3}{\rmd \tau_U}=-\tau_2 e_2,
\end{eqnarray}
where $e_0$ is given by (\ref{circolare}), $e_1$ is given by (\ref{e1}), $e_2=\rmd U/\rmd\alpha\equiv E_{\hat \phi}$, $e_3=-\rmd e_1/\rmd\chi$ and
\beq
\tau_1=-\frac12 \frac{\rmd \kappa}{\rmd \alpha}=-\frac{1}{2\gamma_n^2} \frac{\rmd \kappa}{\rmd \nu}, \,\qquad\, 
\tau_2=-\frac12 \kappa\frac{\rmd  \chi}{\rmd \alpha}=-\frac{\kappa }{2\gamma_n^2} \frac{\rmd \chi}{\rmd \nu}\ .
\eeq
Using (\ref{kappa}) and (\ref{chilie}) the latters become
\begin{eqnarray}
\tau_1&=&-\frac{\nu\gamma_n^4}{\kappa}\left[\frac{k_{\rm (lie)}{}_{\hat \rho}^2}{\gamma_\pm^2} (\nu^2-\nu_\pm^2)+ 2k_{\rm (lie)}{}_{\hat z}^2\,(1+\nu^2)\right], \nonumber \\
\tau_2&=&
\nu\gamma_n^2\frac{k_{\rm (lie)}{}_{\hat \rho} k_{\rm (lie)}{}_{\hat z}}{\kappa} (1+\nu^2_\pm)\ ,
\end{eqnarray}
with $\gamma_\pm=1/\sqrt{1-\nu_\pm^2}$. 
The dual of the chosen FS frame $\{\omega^0, \omega^1,\omega^2,\omega^3\}$ in terms of the frame (\ref{of_zamo}) is given by
\begin{eqnarray}
\label{dualfs}
&&\omega^0=-U^\flat\ , \quad \omega^1=\cos \chi \omega^{\hat \rho}+\sin \chi \omega^{\hat z}\ , \nonumber\\
&&\omega^2\equiv E_{\hat \phi}^\flat=\gamma_n[-\nu\omega^{\hat t}+\omega^{\hat \phi}]\ , \quad \omega^3=\sin \chi \omega^{\hat \rho}-\cos \chi \omega^{\hat z}\ .
\end{eqnarray}
We notice also that the spatial FS frame $e_1,e_2,e_3$ rotates with respect to a Fermi-Walker transported frame along $U$ with angular velocity
\beq
\omega_{\rm (FS)}=\tau_1 e_3 + \tau_2 e_1\ ,
\eeq
which has magnitude
\beq
||\omega_{\rm (FS)}||= |\nu |\gamma_n^2\left[ \frac{k_{\rm (lie)}{}_{\hat \rho}^2}{\gamma_\pm^4} + 4k_{\rm (lie)}{}_{\hat z}^2 \right]^{1/2} .
\eeq
A discussion of special orbits and their FS characterization can be found in \cite{bgijmpd}.

\section{Spinning test particles}

The Mathisson-Papapetrou equations of motion for a spinning test particle are given by
\begin{eqnarray}
\label{papcoreqs1}
\frac{DP^{\mu}}{\rmd \tau_U}&=&-\frac12R^{\mu}{}_{\nu\alpha\beta}U^{\nu}S^{\alpha\beta}\equiv F^{\rm (spin)}{}^{\mu}\\
\label{papcoreqs2}
\frac{DS^{\mu\nu}}{\rmd \tau_U}&=&P^{\mu}U^{\nu}-P^{\nu}U^{\mu}\ ,
\end{eqnarray}
where $P^{\mu}$ is the total 4-momentum of the particle, and $S^{\mu\nu}$ is a (antisymmetric) spin tensor; $U$ is the timelike unit tangent vector of the \lq\lq center line'' used to make the multipole reduction. 
Equations~(\ref{papcoreqs1}) and (\ref{papcoreqs2}) define 
the evolution of $P$ and $S$ only along the world line of  $U$, so a correct interpretation of $U$ is that of being tangent to the  {\it true} 
world-line of the spinning particle. 
In this case, both  $U$ and $P$ are linear combinations of Killing vector fields; this property 
leads to a big simplification since all the FS intrinsic quantities of the world line $U$ (curvature and torsions), the magnitude of $P$, 
the FS frame components and the algebraic invariant of $S$, namely $s^2=\frac12S_{\mu\nu}S^{\mu\nu}$, and other kinematically relevant quantities, will be (covariantly) constant along $U$.

Following the analysis made in \cite{bdfg1,bdfg2}, contracting both sides of equation~(\ref{papcoreqs2}) with $U_\nu$, one obtains
\begin{equation}
\label{Ps}
P^{\mu}=-(U\cdot P)U^\mu -U_\nu \frac{DS^{\mu\nu}}{\rmd \tau_U}\equiv
mU^\mu +P_s^\mu\ ,
\end{equation}
where $m$ is the particle's bare mass \cite{bdfg1,bdfg2}.
Equation~(\ref{papcoreqs2}) implies
\begin{equation}\fl\quad
\label{spinconds}
S_{\hat t\hat \phi}=0\ , \qquad 
S_{\hat \rho\hat z}=0\ , \qquad 
k_{\rm (lie)}{}_{\hat \rho} [\nu_\pm^2S_{{\hat z}{\hat t}}+\nu S_{{\hat z}{\hat \phi}}]+k_{\rm (lie)}{}_{\hat z} [S_{{\hat \rho}{\hat t}} - \nu S_{{\hat \rho}{\hat \phi}}]=0\ .
\end{equation}
The spin tensor then takes the form
\begin{equation}
\label{Sform}
S=\omega^{\hat \rho}\wedge [S_{\hat \rho\hat t}\omega^{\hat t}+S_{\hat \rho\hat \phi}\omega^{\hat \phi}]+\omega^{\hat z}\wedge [S_{\hat z\hat t}\omega^{\hat t}+S_{\hat z\hat \phi}\omega^{\hat \phi}]\ .
\end{equation}
It is clear from (\ref{Ps}) that $P_s$ is orthogonal to $U$; moreover it turns out to be also aligned with $E_{\hat \phi}$  
\begin{equation}
\label{ps}
P_s=m_s E_{\hat \phi}\ ,
\end{equation}
where $m_s\equiv||P_s||$ is given by
\begin{eqnarray}
\label{msdef}
m_s=\gamma_n\{k_{\rm (lie)}{}_{\hat \rho} [\nu S_{{\hat \rho}{\hat t}}+\nu_\pm^2 S_{{\hat \rho}{\hat \phi}}]+k_{\rm (lie)}{}_{\hat z} [\nu S_{{\hat z}{\hat t}} - S_{{\hat z}{\hat \phi}}]\}\ .
\end{eqnarray}
From (\ref{Ps}) and (\ref{ps}) the total 4-momentum $P$  can be written in the form $P=\mu \, U_p$, with
\begin{equation}
\label{Ptot}
U_p=\gamma_p\, [e_{\hat t}+\nu_p e_{\hat \phi}]\ , \quad \nu_p=\frac{\nu+m_s/m}{1+\nu m_s/m}\ ,\quad \mu=\frac{\gamma_n}{\gamma_p}(m+\nu m_s)\ 
\end{equation}
and $\gamma_p=(1-\nu_p^2)^{-1/2}$.
Since $U_p$ is a unit vector, the quantity $\mu$ can be interpreted as the total mass of the particle in the rest-frame of $U_p$.

Let us now consider the equation of motion (\ref{papcoreqs1}). 
The spin-force is equal to:
\begin{eqnarray}\fl\quad
\label{Fspin}
F^{\rm (spin)}&=&\gamma_n\Bigg\{
S_{{\hat \rho}{\hat t}}\left[\partial_{\hat \rho}k_{\rm (lie)}{}_{\hat \rho}-k_{\rm (lie)}{}_{\hat \rho}^2(1+\nu_\pm^2)+\frac{2\nu_\pm^2}{1+\nu_\pm^2}k_{\rm (lie)}{}_{\hat z}^2\right]+\nu  S_{{\hat \rho}{\hat \phi}}\bigg[-\partial_{\hat \rho}k_{\rm (lie)}{}_{\hat \rho}\nonumber\\
\fl\quad
&&+k_{\rm (lie)}{}_{\hat \rho}^2+\frac{k_{\rm (lie)}{}_{\hat z}^2}{\gamma_\pm^2}\bigg]-(S_{{\hat z}{\hat t}}
-\nu S_{{\hat z}{\hat \phi}})\bigg[-\partial_{\hat \rho}k_{\rm (lie)}{}_{\hat z}+\rho \kappa_{\rm (lie)}^2 k_{\rm (lie)}{}_{\hat z}\nonumber\\
\fl\quad
&&+\frac{k_{\rm (lie)}{}_{\hat z}}{k_{\rm (lie)}{}_{\hat \rho}}\frac{\kappa_{\rm (lie)}^2+2k_{\rm (lie)}{}_{\hat \rho}^2\nu_\pm^2}{1+\nu_\pm^2}\bigg]
\Bigg\}e_{\hat \rho}\nonumber\\
\fl\quad
&&+\gamma_n\Bigg\{
(S_{{\hat \rho}{\hat t}}-\nu S_{{\hat \rho}{\hat \phi}})\left[\partial_{\hat z}k_{\rm (lie)}{}_{\hat \rho}+\frac{k_{\rm (lie)}{}_{\hat z}}{k_{\rm (lie)}{}_{\hat \rho}}\frac{\kappa_{\rm (lie)}^2+2k_{\rm (lie)}{}_{\hat \rho}^2\nu_\pm^2}{1+\nu_\pm^2}\right]+S_{{\hat z}{\hat t}}\Bigg[\partial_{\hat z}k_{\rm (lie)}{}_{\hat z}\nonumber\\
\fl\quad
&&-\frac{k_{\rm (lie)}{}_{\hat \rho}^2\nu_\pm^4-k_{\rm (lie)}{}_{\hat z}^2}{1+\nu_\pm^2}\Bigg]-\nu S_{{\hat z}{\hat \phi}}\Bigg[\partial_{\hat z}k_{\rm (lie)}{}_{\hat z}+\frac{(k_{\rm (lie)}{}_{\hat \rho}^2-k_{\rm (lie)}{}_{\hat z}^2)\nu_\pm^2}{1+\nu_\pm^2}\Bigg]
\Bigg\}e_{\hat z}\ ,
\end{eqnarray}
while the term on the left hand side of equation~(\ref{papcoreqs1}) can be written, from (\ref{Ps}) and (\ref{ps}),  as 
\beq
\label{motrad}
\frac{DP}{\rmd \tau_U}=m a(U)+m_s \frac{DE_{\hat \phi}}{\rmd \tau_U}\ ,
\eeq
where $a(U)=\kappa e_1$ and $DE_{\hat \phi}/{\rmd \tau_U}=-\tau_1 e_1+\tau_2 e_3$ are given in (\ref{FSeqs}), and the quantities  $\mu, m, m_s$ are constant along the world line of $U$. 
Hence equation~(\ref{papcoreqs1}) can be written as \footnote{Equivalently, in the FS frame, one obtains:
$\quad 0=m\kappa -m_s \tau_1-F^{\rm (spin)}_1, \quad
0=m_s \tau_2-F^{\rm (spin)}_2\ .$
}
\begin{eqnarray}
\label{eqmoto}
0&=&(m\kappa -m_s\tau_1)\cos\chi+m_s\tau_2\sin\chi-F^{\rm (spin)}_{\hat \rho}\ ,\nonumber\\
0&=&(m\kappa -m_s\tau_1)\sin\chi-m_s\tau_2\cos\chi-F^{\rm (spin)}_{\hat z}\ ,
\end{eqnarray}
or, more explicitly,
\begin{eqnarray}
\label{eqmotonew}
0&=&\gamma_n^2\left[m(\nu^2-\nu_\pm^2)+m_s\frac{\nu}{\gamma_\pm^2}\right]k_{\rm (lie)}{}_{\hat \rho}-F^{\rm (spin)}_{\hat \rho}\ ,\nonumber\\
0&=&\gamma_n^2\left[m(1+\nu^2)+2m_s\nu\right]k_{\rm (lie)}{}_{\hat z}-F^{\rm (spin)}_{\hat z}\ .
\end{eqnarray}

It is useful to introduce the quadratic invariant 
\beq
\label{spininvar}
s^2=\frac12 S_{\mu\nu}S^{\mu\nu}=-S_{\hat \rho\hat t }^2-S_{\hat z\hat t }^2+S_{\hat \rho \hat \phi}^2+S_{\hat z\hat \phi}^2\ . 
\eeq
From equation~(\ref{Sform}) and by using the relations 
\begin{eqnarray}
\omega^{\hat t}&=&\gamma_n [-U^{\flat}+\nu E_{\hat \phi}^\flat]\ , \qquad  \omega^{\hat \rho}=\cos\chi\omega^1+\sin\chi\omega^3\ , \nonumber\\
\omega^{\hat \phi}&=& \gamma_n [-\nu U^{\flat} +E_{\hat \phi}^\flat], \qquad \omega^{\hat z}=\sin\chi\omega^1-\cos\chi\omega^3\ , 
\end{eqnarray}
where $\omega^1$ and $\omega^3$ are defined in equation (\ref{dualfs}) and $X^{\flat}$ denotes the 1-form associated to a vector $X$,
one has the relation
\begin{eqnarray}
S&=&\gamma_n \Big[(S_{\hat \rho\hat t} + \nu S_{\hat \rho\hat \phi} )  U^{\flat}\wedge \omega^{\hat \rho} +(\nu S_{\hat \rho\hat t} + S_{\hat \rho\hat \phi} )\omega^{\hat \rho}\wedge E_{\hat \phi}^\flat\nonumber\\
&&+(S_{\hat z\hat t} + \nu S_{\hat z\hat \phi} )  U^{\flat}\wedge \omega^{\hat z} +(\nu S_{\hat z\hat t} + S_{\hat z\hat \phi} )\omega^{\hat z}\wedge E_{\hat \phi}^\flat \Big]\ .
\end{eqnarray}
Since the components of $S$ are assumed to be constant along $U$, then from the FS formalism one finds
\beq
\frac{DS}{\rmd \tau_U}=m_s E_{\hat \phi}^\flat \wedge U^{\flat}\ ,  
\eeq
or, from equations~(\ref{papcoreqs2}) and (\ref{ps}),
\beq
P_s=m_s E_{\hat \phi}^\flat\ ,
\eeq
with 
\begin{eqnarray}\fl\quad
m_s&=&-\gamma_n \Big\{[(\tau_1+\kappa \nu )\cos\chi-\tau_2\sin\chi]S_{\hat \rho\hat t}+[(\nu \tau_1+\kappa )\cos\chi-\nu\tau_2\sin\chi]S_{\hat \rho\hat \phi}\nonumber\\
\fl\quad
&&+[\tau_2\cos\chi +(\tau_1+\kappa \nu )\sin\chi]S_{\hat z\hat t}+[\nu\tau_2\cos\chi+(\nu \tau_1+\kappa )\sin\chi]S_{\hat z\hat \phi}\Big\}\ .
\end{eqnarray}

To discuss the features of the motion we need to supplement equation~(\ref{eqmoto}) with further conditions. 
We shall do this in the next section following the standard approaches existing in the literature:
\begin{itemize}
\item[1.]
Corinaldesi-Papapetrou \cite{cori51} conditions (CP): $S^{t\nu}=0$,
\item[2.]
Pirani \cite{pir56} conditions (P): $S^{\mu\nu}U_\nu=0$, 
\item[3.]
Tulczyjew \cite{tulc59} conditions (T): $S^{\mu\nu}P_\nu=0$.
\end{itemize}

The above supplementary conditions are necessary but somewhat arbitrary. Aim of our analysis is also that of comparing the physical implications on the motion of a spinning body by each of those conditions and identify the most significant one. 

Since our general analysis leads to rather general complicated expressions we shall specify the general equations to the
particular case of hyperplanes characterized by $k_{\rm (lie)}{}_{\hat z}=0$ (and so $\partial_{\hat \rho}k_{\rm (lie)}{}_{\hat z}=0=\partial_{\hat z}k_{\rm (lie)}{}_{\hat \rho}$ too), namely confining ourselves to the mirror symmetry hyperplanes such that $\psi_{,z}=0$, which exist for many interesting solutions.
In this case, the circular orbit with $\nu=\nu_\pm$ ($\zeta=\zeta_\pm$) are geodesics. Now
the constraints (\ref{spinconds}) become
\begin{equation}
\label{spincondskzeq0}
S_{\hat t\hat \phi}=0\ , \qquad 
S_{\hat \rho\hat z}=0\ , \qquad 
S_{{\hat t}{\hat z}}=\frac{\nu}{\nu_\pm^2} S_{{\hat z}{\hat \phi}}\ ;
\end{equation}
the quantity $m_s$ defined by (\ref{msdef}) writes as 
\begin{eqnarray}
\label{msdefkzeq0}
m_s=\gamma_n k_{\rm (lie)}{}_{\hat \rho} [\nu S_{{\hat \rho}{\hat t}}+\nu_\pm^2 S_{{\hat \rho}{\hat \phi}}]\ ,
\end{eqnarray}
and the spin force (\ref{Fspin}) simplifies as follows:
\begin{eqnarray}\fl\quad
\label{Fspinkzeq0}
F^{\rm (spin)}&&=\gamma_n\Bigg\{
S_{{\hat \rho}{\hat t}}\left[\partial_{\hat \rho}k_{\rm (lie)}{}_{\hat \rho}-k_{\rm (lie)}{}_{\hat \rho}^2(1+\nu_\pm^2)\right]+\nu  S_{{\hat \rho}{\hat \phi}}\bigg[-\partial_{\hat \rho}k_{\rm (lie)}{}_{\hat \rho}+k_{\rm (lie)}{}_{\hat \rho}^2\bigg]
\Bigg\}e_{\hat \rho}\nonumber\\
\fl\quad
&&+\gamma_n\Bigg\{
S_{{\hat z}{\hat t}}\left[\partial_{\hat z}k_{\rm (lie)}{}_{\hat z}-\frac{k_{\rm (lie)}{}_{\hat \rho}^2\nu_\pm^4}{1+\nu_\pm^2}\right]-\nu S_{{\hat z}{\hat \phi}}\Bigg[\partial_{\hat z}k_{\rm (lie)}{}_{\hat z}+\frac{k_{\rm (lie)}{}_{\hat \rho}^2\nu_\pm^2}{1+\nu_\pm^2}\Bigg]
\Bigg\}e_{\hat z}\ .
\end{eqnarray}
Recalling the conditions (\ref{spincondskzeq0}), the equations of motion (\ref{eqmotonew}) reduce to
\begin{eqnarray}\fl\quad
\label{eqmotokzeq0}
0&=&\gamma_n\left[m(\nu^2-\nu_\pm^2)+m_s\frac{\nu}{\gamma_\pm^2}\right]k_{\rm (lie)}{}_{\hat \rho}
-\bigg\{S_{{\hat \rho}{\hat t}}\left[\partial_{\hat \rho}k_{\rm (lie)}{}_{\hat \rho}-k_{\rm (lie)}{}_{\hat \rho}^2(1+\nu_\pm^2)\right]\nonumber\\
\fl\quad
&&+\nu  S_{{\hat \rho}{\hat \phi}}\bigg[-\partial_{\hat \rho}k_{\rm (lie)}{}_{\hat \rho}+k_{\rm (lie)}{}_{\hat \rho}^2\bigg]\bigg\}\nonumber\\
\fl\quad
0&=&\gamma_n \nu S_{{\hat z}{\hat \phi}} \partial_{\hat z}k_{\rm (lie)}{}_{\hat z}\left[1+\frac1{\nu_\pm^2}\right]\ .
\end{eqnarray}
The latter equation implies $S_{{\hat z}{\hat \phi}}=0$, since $\partial_{\hat z}k_{\rm (lie)}{}_{\hat z}\not=0$ in general, and so $S_{{\hat t}{\hat z}}=0$ too. 
Therefore, the spin tensor turns out to be completely determined in this case by two components only, namely 
$S_{{\hat t}{\hat \rho}}$ and $S_{{\hat \rho}{\hat \phi}}$, related by the first equation of (\ref{eqmotokzeq0}).

\subsection{The Corinaldesi-Papapetrou (CP) supplementary conditions}

The CP supplementary conditions require $S_{\hat t \hat \rho}=0=S_{\hat t \hat z}$, so that 
\beq
\label{SdefCP}
S={\mathcal S}_{(CP)}\wedge \omega^{\hat \phi}\ , \qquad 
{\mathcal S}_{(CP)}=S_{{\hat \rho}{\hat \phi}}\omega^{\hat \rho}+S_{{\hat z}{\hat \phi}}\omega^{\hat z} .
\eeq
From equation~(\ref{msdef}) we have that 
\beq
m_s=\frac{s\gamma_n}{\kappa_{\rm (lie)}}[k_{\rm (lie)}{}_{\hat \rho}^2\nu_\pm^2-k_{\rm (lie)}{}_{\hat z}^2]\ ;
\eeq
the spin force (see equation~(\ref{Fspin})) is given by 
\beq
\label{fspincp1}
F^{(\rm spin)}=s\gamma_n \nu \mathcal{F}^{\rm (Riem)}\ ,
\eeq
where $\mathcal{F}^{\rm (Riem)}$ does not depend on $s$ and $\nu$:
\begin{eqnarray}
\label{fspincp2}
\fl\quad
\mathcal{F}^{\rm (Riem)}&=&\frac1{\kappa_{\rm (lie)}}\left\{
-\frac12\partial_{\hat \rho}\kappa_{\rm (lie)}^2+\kappa_{\rm (lie)}^2k_{\rm (lie)}{}_{\hat \rho}+\frac{\kappa_{\rm (lie)}^2k_{\rm (lie)}{}_{\hat z}^2}{k_{\rm (lie)}{}_{\hat \rho}}\left[\rho k_{\rm (lie)}{}_{\hat \rho}+\frac1{1+\nu_\pm^2}\right]
\right\}e_{\hat \rho} \nonumber\\
\fl\quad
&&+\frac1{\kappa_{\rm (lie)}}\left\{
-\frac12\partial_{\hat z}\kappa_{\rm (lie)}^2+\kappa_{\rm (lie)}^2k_{\rm (lie)}{}_{\hat z}
\right\}e_{\hat z}\ .
\end{eqnarray}
Now, by solving the first equation of (\ref{eqmotonew}) with respect to $s$ once equations (\ref{fspincp1}) and (\ref{fspincp2}) have been used, we get
\beq
\label{ssolCP}
s=-\frac{m\gamma_n}{\nu}(\nu^2-\nu_\pm^2)
\left[\frac{\gamma_n^2}{\gamma_\pm^2}\frac{k_{\rm (lie)}{}_{\hat \rho}^2\nu_\pm^2-k_{\rm (lie)}{}_{\hat z}^2}{\kappa_{\rm (lie)}}-\frac{\mathcal{F}^{\rm (Riem)}_{\hat \rho}}{k_{\rm (lie)}{}_{\hat \rho}}\right]^{-1}
\ ,
\eeq
and the corresponding solution for $\nu \equiv {}^s\nu_{\pm}$ is obtained by substituting equation~(\ref{ssolCP}) into the second equation of (\ref{eqmotonew}): 
\begin{eqnarray}\fl\quad
{}^s\nu_{\pm}&&= \pm \Bigg\{ 
-1+
k_{\rm (lie)}{}_{\hat \rho}(1+\nu_\pm^2)\left\{\partial_{\hat z}\kappa_{\rm (lie)}^2-2k_{\rm (lie)}{}_{\hat z}\left[\frac{k_{\rm (lie)}{}_{\hat \rho}^2}{\gamma_\pm^2}+2k_{\rm (lie)}{}_{\hat z}^2\right]\right\}\cdot\nonumber\\
\fl\quad
&&\cdot\bigg\{k_{\rm (lie)}{}_{\hat \rho}\partial_{\hat z}\kappa_{\rm (lie)}^2-k_{\rm (lie)}{}_{\hat z}\partial_{\hat \rho}\kappa_{\rm (lie)}^2+2\frac{\kappa_{\rm (lie)}^2 k_{\rm (lie)}{}_{\hat z}^3}{k_{\rm (lie)}{}_{\hat \rho}}\left[\rho k_{\rm (lie)}{}_{\hat \rho}+\frac1{1+\nu_\pm^2}\right]\bigg\}^{-1}
\Bigg\}^{1/2}\ .
\end{eqnarray}
Obviously, by introducing this value of $\nu$ into the previous equation~(\ref{ssolCP}), we get the solution $s$ as a function of $\rho$ and $z$.

As anticipated, let us now consider  the case of circular orbits on mirror symmetry hyperplanes characterized by $k_{\rm (lie)}{}_{\hat z}=0$. The spin tensor writes as
\beq
\label{SdefCPkzeq0}
S=s \, \omega^{\hat \rho}\wedge \omega^{\hat \phi}\ ,
\eeq
being ${\mathcal S}_{(CP)}=s\,\omega^{\hat \rho}$ as from equations~(\ref{spinconds}), (\ref{spininvar}) and (\ref{SdefCP}), and the first of equations~(\ref{eqmotokzeq0}) reduces to
\begin{equation}
\label{eqmotoCP}
0=m(\nu^2-\nu_\pm^2)+\frac{s\nu}{\gamma_n}\left\{\partial_{\hat \rho}\ln{k_{\rm (lie)}{}_{\hat \rho}}-k_{\rm (lie)}{}_{\hat \rho}\left[1-\frac{\gamma_n^2}{\gamma_\pm^2}\nu_\pm^2\right]\right\}\ .
\end{equation}
By solving this equation with respect to $s$, we obtain
\begin{equation}
\label{solCPkzeq0}
s=m\frac{\gamma_n}{\nu}\frac{\nu^2-\nu_\pm^2}{-\partial_{\hat \rho}\ln{k_{\rm (lie)}{}_{\hat \rho}}+k_{\rm (lie)}{}_{\hat \rho}\left[1-\frac{\gamma_n^2}{\gamma_\pm^2}\nu_\pm^2\right]}\ .
\end{equation}
In the limit of small $s$, the preceding expression leads to
\begin{equation}\fl\quad
\label{solCPexpnu}
\nu= \pm\nu_\pm+{\mathcal N}^{(CP)}s+O(s^2)\ , \qquad {\mathcal N}^{(CP)}=\frac1{2\gamma_\pm}\left[-\partial_{\hat \rho}\ln{k_{\rm (lie)}{}_{\hat \rho}}+\frac{k_{\rm (lie)}{}_{\hat \rho}}{\gamma_\pm^2}\right]\ .
\end{equation}
The corresponding angular velocity $\zeta$ and its reciprocal are
\begin{equation}\fl\quad
\label{zetaCP}
\zeta= \pm\zeta_\pm +\frac{e^{2\psi}}{\rho}{\mathcal N}^{(CP)}s+O(s^2)\ , \qquad \frac1{\zeta}= \pm\frac{1}{\zeta_\pm}-\frac{e^{2\psi}}{\rho\zeta_\pm^2}{\mathcal N}^{(CP)}s+O(s^2)\ ,
\end{equation}
the metric function $\psi$ being evaluated on the symmetry hyperplane under consideration.
The total 4-momentum $P$ is given by equation~(\ref{Ptot}) with
\begin{equation}
m_s=s\gamma_n k_{\rm (lie)}{}_{\hat \rho}\nu_\pm^2\ .
\end{equation}
In the limit of small $s$ the linear velocity $\nu_p$ reduces to
\begin{equation}\fl\quad
\label{solCPexpnup}
\nu_p=\pm\nu_\pm+{\mathcal N}_p^{(CP)}s+O(s^2)\ , \qquad {\mathcal N}_p^{(CP)}= {\mathcal N}^{(CP)}+\frac{k_{\rm (lie)}{}_{\hat \rho}}m\frac{\nu_\pm^2}{\gamma_\pm}\ .
\end{equation}
The corresponding angular velocity $\zeta_p$ and its reciprocal are 
\begin{equation}\fl\quad
\label{zetapCP}
\zeta_p= \pm\zeta_\pm +\frac{e^{2\psi}}{\rho}{\mathcal N}_p^{(CP)}s+O(s^2)\ , \qquad \frac1{\zeta_p}= \pm\frac{1}{\zeta_\pm}-\frac{e^{2\psi}}{\rho\zeta_\pm^2}{\mathcal N}_p^{(CP)}s+O(s^2)\ .
\end{equation}

\subsection{The Pirani (P) supplementary conditions}

The P supplementary conditions ($S^{\mu\nu}U_\nu=0$) require 
\beq
S_{\hat \rho \hat t}+S_{\hat \rho \hat \phi}\nu=0\ , \qquad
S_{\hat z\hat t}+S_{\hat z\hat \phi}\nu=0\ ,
\eeq
so that
\beq
\label{SdefP}
S={\mathcal S}_{(P)}\wedge E_{\hat \phi}^\flat\ , \qquad {\mathcal S}_{(P)}=
\frac{1}{\gamma_n}[S_{{\hat \rho}{\hat \phi}}\omega^{\hat \rho}+S_{{\hat z}{\hat \phi}}\omega^{\hat z}]\ .
\eeq 
From equation~(\ref{msdef}) we have that 
\beq\fl\quad
m_s=-s\frac{\gamma_n^2}{\gamma_\pm^2\Omega}[k_{\rm (lie)}{}_{\hat \rho}^2(\nu^2-\nu_\pm^2)+ 2\gamma_\pm^2 k_{\rm (lie)}{}_{\hat z}^2\,(1+\nu^2)]
\eeq
with
\beq
\Omega=\left[ \frac{k_{\rm (lie)}{}_{\hat \rho}^2}{\gamma_\pm^4} + 4k_{\rm (lie)}{}_{\hat z}^2 \right]^{1/2}\ ;
\eeq
the spin force (see equation~(\ref{Fspin})) is given by 
\beq
\label{fspinp1}
F^{(\rm spin)}=s\gamma_n^2 \nu \mathcal{F}^{\rm (Riem)}\ ,
\eeq
where $\mathcal{F}^{\rm (Riem)}$ does not depend on $s$ and $\nu$; then
\begin{eqnarray}
\label{fspinp2}
\fl\quad
\mathcal{F}^{\rm (Riem)}&=&\frac{2}{\Omega}\Bigg\{
-\frac1{\gamma_\pm^2}k_{\rm (lie)}{}_{\hat \rho}\partial_{\hat \rho}k_{\rm (lie)}{}_{\hat \rho}-2k_{\rm (lie)}{}_{\hat z}\partial_{\hat \rho}k_{\rm (lie)}{}_{\hat z}
+2\rho\kappa_{\rm (lie)}^2k_{\rm (lie)}{}_{\hat z}^2\nonumber\\
\fl\quad
&&+\frac{2+\nu_\pm^2}{2\gamma_\pm^2}k_{\rm (lie)}{}_{\hat \rho}^3 + \frac{5+\nu_\pm^2(4+3\nu_\pm^2)}{2(1+\nu_\pm^2)}k_{\rm (lie)}{}_{\hat \rho}k_{\rm (lie)}{}_{\hat z}^2+\frac2{1+\nu_\pm^2}\frac{k_{\rm (lie)}{}_{\hat z}^4}{k_{\rm (lie)}{}_{\hat \rho}}
\Bigg\}e_{\hat \rho} \nonumber\\
\fl\quad
&&+\frac{2}{\Omega}\Bigg\{
-\frac1{\gamma_\pm^2}k_{\rm (lie)}{}_{\hat \rho}\partial_{\hat z}k_{\rm (lie)}{}_{\hat \rho}-2k_{\rm (lie)}{}_{\hat z}\partial_{\hat z}k_{\rm (lie)}{}_{\hat z}\nonumber\\
\fl\quad
&&+\frac1{\gamma_\pm^2}k_{\rm (lie)}{}_{\hat \rho}^2 k_{\rm (lie)}{}_{\hat z}
\Bigg\}e_{\hat z}\ .
\end{eqnarray}
Now, by solving the first equation of (\ref{eqmotonew}) with respect to $s$ once equations (\ref{fspinp1}) and (\ref{fspinp2}) have been used, we obtain
\beq
\label{ssolP}
s=-m\gamma_n^2 (\nu^2-\nu_\pm^2)\left[\frac{1}{\gamma_\pm^2}\frac{\kappa\tau_1}{\Omega}-\gamma_n^2\nu\frac{\mathcal{F}^{\rm (Riem)}_{\hat \rho}}{k_{\rm (lie)}{}_{\hat \rho}}\right]^{-1}
\ ,
\eeq
and the corresponding solution for $\nu \equiv {}^s\nu_{\pm}$ is derived by substituting equation~(\ref{ssolP}) into the second equation of (\ref{eqmotonew}): 
\begin{eqnarray}\fl\quad
{}^s\nu_{\pm}&=& \pm \Bigg\{ 
-1+
\frac{k_{\rm (lie)}{}_{\hat \rho}(1+\nu_\pm^2)}{\mathcal D}\bigg[
-4k_{\rm (lie)}{}_{\hat z}\partial_{\hat z}k_{\rm (lie)}{}_{\hat z}-\frac2{\gamma_\pm^2}k_{\rm (lie)}{}_{\hat \rho}\partial_{\hat z}k_{\rm (lie)}{}_{\hat \rho}\nonumber\\
\fl\quad
&&+\frac1{\gamma_\pm^4}k_{\rm (lie)}{}_{\hat \rho}^2k_{\rm (lie)}{}_{\hat z}
\bigg]\Bigg\}^{1/2}\nonumber\\
\fl\quad
{\mathcal D}&=&2\left[-\frac{k_{\rm (lie)}{}_{\hat \rho}^2}{\gamma_\pm^2}+2k_{\rm (lie)}{}_{\hat z}^2\right]\partial_{\hat z}k_{\rm (lie)}{}_{\hat \rho}+2k_{\rm (lie)}{}_{\hat \rho}k_{\rm (lie)}{}_{\hat z}(\nu_\pm^2-3)\partial_{\hat z}k_{\rm (lie)}{}_{\hat z}\nonumber\\
\fl\quad
&&+\frac1{1+\nu_\pm^2}\frac{k_{\rm (lie)}{}_{\hat z}}{k_{\rm (lie)}{}_{\hat \rho}}\left[\frac{k_{\rm (lie)}{}_{\hat \rho}^4}{\gamma_\pm^4}-(3\nu_\pm^2+7)(1+\nu_\pm^2)k_{\rm (lie)}{}_{\hat \rho}^2k_{\rm (lie)}{}_{\hat z}^2-4k_{\rm (lie)}{}_{\hat z}^4\right]\ .
\end{eqnarray}

Let us consider now the case $k_{\rm (lie)}{}_{\hat z}=0$. The spin tensor writes as
\beq
\label{SdefPkzeq0}
S=s \, \omega^{\hat \rho}\wedge E_{\hat \phi}^\flat\ ,
\eeq
being ${\mathcal S}_{(P)}=s\,\omega^{\hat \rho}$ as from equations~(\ref{spinconds}), (\ref{spininvar}) and (\ref{SdefP}), so that $(S_{\hat \rho \hat t}, S_{\hat \rho \hat \phi})=(-s\gamma_n\nu, s\gamma_n)$ and equation~(\ref{eqmotokzeq0}) reduces to
\begin{equation}\fl\quad
\label{eqmotoP}
0=m(\nu^2-\nu_\pm^2)+s\nu\left\{2\partial_{\hat \rho}\ln{k_{\rm (lie)}{}_{\hat \rho}}-k_{\rm (lie)}{}_{\hat \rho}\left[2+\nu_\pm^2+\frac{\gamma_n^2}{\gamma_\pm^2}(\nu^2-\nu_\pm^2)\right]\right\}\ .
\end{equation}
By solving this equation with respect to $s$, we obtain
\begin{equation}
\label{solPkzeq0}
s=\frac{m}{\nu}\frac{\nu^2-\nu_\pm^2}{-2\partial_{\hat \rho}\ln{k_{\rm (lie)}{}_{\hat \rho}}+k_{\rm (lie)}{}_{\hat \rho}\left[2+\nu_\pm^2+\frac{\gamma_n^2}{\gamma_\pm^2}(\nu^2-\nu_\pm^2)\right]}\ .
\end{equation}
In the limit of small $s$ the preceding expression leads to
\begin{equation}\fl\quad
\label{solPexpnu}
\nu= \pm\nu_\pm+{\mathcal N}^{(P)}s+O(s^2)\ , \quad {\mathcal N}^{(P)}=-\frac1{2m}\left[2\partial_{\hat \rho}\ln{k_{\rm (lie)}{}_{\hat \rho}}-k_{\rm (lie)}{}_{\hat \rho}(2+\nu_\pm^2)\right]\ .
\end{equation}
The corresponding angular velocity $\zeta$ and its reciprocal are
\begin{equation}\fl\quad
\label{zetaP}
\zeta= \pm\zeta_\pm +\frac{e^{2\psi}}{\rho}{\mathcal N}^{(P)}s+O(s^2)\ , \qquad \frac1{\zeta}= \pm\frac{1}{\zeta_\pm}-\frac{e^{2\psi}}{\rho\zeta_\pm^2}{\mathcal N}^{(P)}s+O(s^2)\ .
\end{equation}
The total 4-momentum $P$ is given by equation~(\ref{Ptot}) with
\begin{equation}
m_s=-s\gamma_n^2 k_{\rm (lie)}{}_{\hat \rho}[\nu^2-\nu_\pm^2]\ , \qquad  \nu_p=\nu+O(s^2)\ .
\end{equation}
The  angular velocity $\zeta_p$ and its reciprocal are
\begin{equation}
\label{zetapP}
\zeta_p= \zeta+O(s^2)\ , \qquad \frac1{\zeta_p}= \frac{1}{\zeta}+O(s^2)\ ,
\end{equation}
with $\zeta$ given by equation~(\ref{zetaP}).

\subsection{The Tulczyjew (T) supplementary conditions}

The T supplementary conditions ($S^{\mu\nu}P_\nu=0$) require 
\beq
S_{\hat \rho \hat t}+S_{\hat \rho \hat \phi}\nu_p=0\ , \qquad
S_{\hat z\hat t}+S_{\hat z\hat \phi}\nu_p=0\ ,
\eeq
so that
\beq\fl\quad
\label{SdefT}
S={\mathcal S}_{(T)}\wedge \gamma_p [-\nu_p \omega^{\hat t}+\omega^{\hat \phi}]\ , 
\quad {\mathcal S}_{(T)}=\frac{1}{\gamma_p}[S_{{\hat \rho}{\hat \phi}}\omega^{\hat \rho}+S_{{\hat z}{\hat \phi}}\omega^{\hat z}]\ .
\eeq
From equation~(\ref{msdef}) we have that 
\beq\fl\quad
m_s=-\frac{s\gamma_n\gamma_p}{\Lambda}[k_{\rm (lie)}{}_{\hat \rho}^2(\nu\nu_p-\nu_\pm^2)(\nu-\nu_p\nu_\pm^2)+ k_{\rm (lie)}{}_{\hat z}^2(\nu+\nu_p)(1+\nu\nu_p)]
\eeq
where
\beq
\Lambda=\left[ k_{\rm (lie)}{}_{\hat \rho}^2(\nu-\nu_p\nu_\pm^2)^2 + k_{\rm (lie)}{}_{\hat z}^2 (\nu+\nu_p)^2\right]^{1/2};
\eeq
the spin force (see equation~(\ref{Fspin})) is given by 
\beq
\label{fspint1}
F^{(\rm spin)}=s\gamma_n\gamma_p\mathcal{F}^{\rm (Riem)}\ ,
\eeq
where $\mathcal{F}^{\rm (Riem)}$ does not depend on $s$ (but it depends on $\nu$ in this case):
\begin{eqnarray}
\label{fspint2}
\fl\quad
\mathcal{F}^{\rm (Riem)}&=&-\frac{\nu+\nu_p}{\Lambda}
\Bigg\{
(\nu-\nu_p\nu_\pm^2)k_{\rm (lie)}{}_{\hat \rho}\partial_{\hat \rho}k_{\rm (lie)}{}_{\hat \rho}+(\nu+\nu_p)k_{\rm (lie)}{}_{\hat z}\partial_{\hat \rho}k_{\rm (lie)}{}_{\hat z}\nonumber\\
\fl\quad
&&-\rho\kappa_{\rm (lie)}^2k_{\rm (lie)}{}_{\hat z}^2(\nu+\nu_p)-\frac{\nu-\nu_p\nu_\pm^2}{\nu+\nu_p}[\nu+\nu_p(1+\nu_\pm^2)]k_{\rm (lie)}{}_{\hat \rho}^3 \nonumber\\
\fl\quad
&&-\frac1{\nu+\nu_p}\bigg\{ \nu^2\left[1+\frac1{1+\nu_\pm^2}\right]+\nu\nu_p\left[\nu_\pm^2+\frac2{1+\nu_\pm^2}\right]\nonumber\\
\fl\quad
&&+\nu_p^2\left[2\nu_\pm^2+\frac1{1+\nu_\pm^2}\right] \bigg\}k_{\rm (lie)}{}_{\hat \rho}k_{\rm (lie)}{}_{\hat z}^2-\frac{\nu+\nu_p}{1+\nu_\pm^2}\frac{k_{\rm (lie)}{}_{\hat z}^4}{k_{\rm (lie)}{}_{\hat \rho}}
\Bigg\}e_{\hat \rho} \nonumber\\
\fl\quad
&&-\frac{\nu+\nu_p}{\Lambda}
\Bigg\{
(\nu-\nu_p\nu_\pm^2)k_{\rm (lie)}{}_{\hat \rho}\partial_{\hat z}k_{\rm (lie)}{}_{\hat \rho}+(\nu+\nu_p)k_{\rm (lie)}{}_{\hat z}\partial_{\hat z}k_{\rm (lie)}{}_{\hat z}\nonumber\\
\fl\quad
&& - k_{\rm (lie)}{}_{\hat z}[k_{\rm (lie)}{}_{\hat \rho}^2(\nu-\nu_p\nu_\pm^2) + k_{\rm (lie)}{}_{\hat z}^2 (\nu-\nu_p)]
\Bigg\}e_{\hat z}\ .
\end{eqnarray}
Contrary to the previous cases, the T supplementary conditions imply that $F^{(\rm spin)}$ does not depend linearly on $\nu$. This makes the properties of the circular motion of the spinning particles much richer but also less tractable analytically.
Solving both equations~(\ref{eqmotonew}) with respect to $s$ using (\ref{fspint1}) and (\ref{fspint2}), we find that the following relations hold simultaneously
\begin{eqnarray}
\label{ssolT}
s&=&-m\frac{\gamma_n}{\gamma_p}(\nu^2-\nu_\pm^2)\left[\frac{\gamma_n^2}{\gamma_\pm^2}\nu{\tilde m_s}-\frac{\mathcal{F}^{\rm (Riem)}_{\hat \rho}}{k_{\rm (lie)}{}_{\hat \rho}}\right]^{-1}\nonumber\\
s&=&-m\frac{\gamma_n}{\gamma_p}(1+\nu^2)\left[2\gamma_n^2\nu{\tilde m_s}-\frac{\mathcal{F}^{\rm (Riem)}_{\hat z}}{k_{\rm (lie)}{}_{\hat z}}\right]^{-1}\ ,
\end{eqnarray}
where the quantity $\tilde m_s$ stands for $\tilde m_s=m_s/(s\gamma_n\gamma_p)$. By eliminating $s$, we have that $\nu_p$ must satisfy the following equation:
\beq
\label{eqnupt}
A\nu_p^2+B\nu_p+C=0\ ,
\eeq
with
\begin{eqnarray}\fl\quad
\phantom{+}A&=&[(\nu^2-\nu_\pm^2)\nu_\pm^2 k_{\rm (lie)}{}_{\hat \rho}^2+(1+\nu^2)k_{\rm (lie)}{}_{\hat z}^2]k_{\rm (lie)}{}_{\hat \rho}\partial_{\hat z}k_{\rm (lie)}{}_{\hat \rho}\nonumber\\
\fl\quad
&&+k_{\rm (lie)}{}_{\hat \rho}^2\left[-\frac{\nu^2}{\gamma_\pm^2}+2\nu_\pm^2\right]k_{\rm (lie)}{}_{\hat z}\partial_{\hat z}k_{\rm (lie)}{}_{\hat z}
-k_{\rm (lie)}{}_{\hat z}\bigg[-\nu_\pm^4(2+\nu_\pm^2)\frac{1+\nu^2}{1+\nu_\pm^2}k_{\rm (lie)}{}_{\hat \rho}^4 \nonumber\\
\fl\quad
&&+ (2\nu^2\nu_\pm^2+3\nu^2+1)k_{\rm (lie)}{}_{\hat \rho}^2 k_{\rm (lie)}{}_{\hat z}^2+k_{\rm (lie)}{}_{\hat z}^4\frac{1+\nu^2}{1+\nu_\pm^2}\bigg]
\nonumber\\
\fl\quad
-\frac{B}{\nu}&=&\left[\frac{\nu^2-\nu_\pm^2}{\gamma_\pm^2} k_{\rm (lie)}{}_{\hat \rho}^2-2(1+\nu^2)k_{\rm (lie)}{}_{\hat z}^2\right]k_{\rm (lie)}{}_{\hat \rho}\partial_{\hat z}k_{\rm (lie)}{}_{\hat \rho}\nonumber\\
\fl\quad
&&+k_{\rm (lie)}{}_{\hat \rho}^2\left[3(\nu^2-\nu_\pm^2)-\nu^2\nu_\pm^2+1\right]k_{\rm (lie)}{}_{\hat z}\partial_{\hat z}k_{\rm (lie)}{}_{\hat z}\nonumber\\
\fl\quad
&&+k_{\rm (lie)}{}_{\hat z}\bigg\{\frac{\nu_\pm^2}{1+\nu_\pm^2}\{[3+\nu_\pm^2(2+\nu_\pm^2)]\nu^2+(2+\nu_\pm^2)(1+\nu_\pm^4)\}k_{\rm (lie)}{}_{\hat \rho}^4 \nonumber\\
\fl\quad
&&+(1+\nu^2)(3+\nu_\pm^2)k_{\rm (lie)}{}_{\hat \rho}^2k_{\rm (lie)}{}_{\hat z}^2+2k_{\rm (lie)}{}_{\hat z}^4 \frac{1+\nu^2}{1+\nu_\pm^2}\bigg\}
\nonumber\\
\fl\quad
-\frac{C}{\nu^2}&=&\left[(\nu^2-\nu_\pm^2)k_{\rm (lie)}{}_{\hat \rho}^2-(1+\nu^2)k_{\rm (lie)}{}_{\hat z}^2\right]k_{\rm (lie)}{}_{\hat \rho}\partial_{\hat z}k_{\rm (lie)}{}_{\hat \rho}\nonumber\\
\fl\quad
&&+k_{\rm (lie)}{}_{\hat \rho}^2\left[2\nu^2-\nu_\pm^2+1\right]k_{\rm (lie)}{}_{\hat z}\partial_{\hat z}k_{\rm (lie)}{}_{\hat z}\nonumber\\
\fl\quad
&&+\frac{k_{\rm (lie)}{}_{\hat z}}{1+\nu_\pm^2}\bigg\{\{[\nu_\pm^2-(1+\nu_\pm^2)^2]\nu^2-\nu_\pm^4(2+\nu_\pm^2)\}k_{\rm (lie)}{}_{\hat \rho}^4 \nonumber\\
\fl\quad
&&+(1+\nu_\pm^2)[\nu^2+2\nu_\pm^2+3]k_{\rm (lie)}{}_{\hat \rho}^2k_{\rm (lie)}{}_{\hat z}^2+k_{\rm (lie)}{}_{\hat z}^4 (1+\nu^2)\bigg\}\ .
\end{eqnarray}
Let $\nu_p^{(\pm)}$ the solutions of equation (\ref{eqnupt}). By substituting $\nu_p=\nu_p^{(\pm)}$ into either equation~(\ref{ssolT}), we obtain a relation between $\nu$ and $s$,
which must be considered together with the following further equation directly descending from the definition (\ref{Ptot}) of $\nu_p$:
\beq
\label{ssolTfromnup}
s=-m\frac{\nu-\nu_p}{\gamma_n\gamma_p{\tilde m_s}(1-\nu\nu_p)}\bigg\vert_{\nu_p=\nu_p^{(\pm)}}\ .
\eeq
As a result, solutions for both quantities $\nu$ and $s$ can be derived explicitly. They are very complicated and poorly illuminating, hence
let us consider the case $k_{\rm (lie)}{}_{\hat z}=0$. The spin tensor writes as
\beq
\label{SdefTkzeq0}
S=s \, \omega^{\hat \rho}\wedge 
\gamma_p [-\nu_p \omega^{\hat t}+\omega^{\hat \phi}]\ ,
\eeq
being ${\mathcal S}_{(T)}=s\,\omega^{\hat \rho}$ as from equations~(\ref{spinconds}), (\ref{spininvar}) and (\ref{SdefT}), so that $(S_{\hat \rho \hat t}, S_{\hat \rho \hat \phi})=(-s\gamma_p\nu_p, s\gamma_p)$ and equation~(\ref{eqmotokzeq0}) reduces to
\begin{eqnarray}\fl\quad
\label{eqmotoT}
0&=&m(\nu^2-\nu_\pm^2)+s\frac{\gamma_p}{\gamma_n}\bigg\{(\nu+\nu_p)\partial_{\hat \rho}\ln{k_{\rm (lie)}{}_{\hat \rho}}-k_{\rm (lie)}{}_{\hat \rho}\bigg[\nu+\nu_p(1+\nu_\pm^2)\nonumber\\
\fl\quad
&&+\nu\frac{\gamma_n^2}{\gamma_\pm^2}(\nu\nu_p-\nu_\pm^2)\bigg]\bigg\}\ .
\end{eqnarray}
By solving this equation with respect to $s$, we obtain
\begin{equation}\fl\quad
\label{solTkzeq0}
s=m\frac{\gamma_n}{\gamma_p}\frac{\nu^2-\nu_\pm^2}{-(\nu+\nu_p)\partial_{\hat \rho}\ln{k_{\rm (lie)}{}_{\hat \rho}}+k_{\rm (lie)}{}_{\hat \rho}\left[\nu+\nu_p(1+\nu_\pm^2)+\nu\frac{\gamma_n^2}{\gamma_\pm^2}(\nu\nu_p-\nu_\pm^2)\right]}\ .
\end{equation}
Recalling its definition (\ref{msdefkzeq0}), $m_s$ becomes
\begin{equation}
m_s=-s\gamma_n\gamma_p k_{\rm (lie)}{}_{\hat \rho}[\nu\nu_p-\nu_\pm^2]\ , 
\end{equation}
and using equation~(\ref{Ptot}) for $\nu_p$, we obtain 
\beq
\label{sfromms}
s=\frac{m}{\gamma_n\gamma_p}\frac1{k_{\rm (lie)}{}_{\hat \rho}}\frac{\nu-\nu_p}{(1-\nu\nu_p)(\nu\nu_p-\nu_\pm^2)}\ ; 
\eeq
this condition  must be considered together with equation~(\ref{solTkzeq0}).
By eliminating $s$ from  equations~(\ref{solTkzeq0}) and (\ref{sfromms}), and solving with respect to $\nu_p$, we have that 
\begin{eqnarray}
\label{nupsol}
\nu_p^{(\pm)}&=&\frac12\frac{3\nu\nu_\pm^2 k_{\rm (lie)}{}_{\hat \rho}\pm\sqrt{\Psi}}{(1+\nu^2+\nu_\pm^2)k_{\rm (lie)}{}_{\hat \rho}-\partial_{\hat \rho}\ln{k_{\rm (lie)}{}_{\hat \rho}}}\nonumber\\
\Psi&=&\left\{2\nu\partial_{\hat \rho}\ln{k_{\rm (lie)}{}_{\hat \rho}}-\frac{k_{\rm (lie)}{}_{\hat \rho}}{\nu}[\nu_\pm^2(\nu^2-\nu_\pm^2)+\nu^2(2+\nu^2)]\right\}^2\nonumber\\
&&-\frac{k_{\rm (lie)}{}_{\hat \rho}^2}{\nu^2}(\nu^2-\nu_\pm^2)^2[\nu^2(\nu^2+4\nu_\pm^2)+\nu_\pm^4]\ .
\end{eqnarray}
By substituting $\nu_p=\nu_p^{(\pm)}$ for instance into equation~(\ref{solTkzeq0}), we obtain a relation between $\nu$ and $s$.
The reality condition of (\ref{nupsol}) requires that $\nu$ takes values outside the interval $({\bar \nu}_-,{\bar \nu}_+)$, 
with 
\begin{eqnarray}
\label{nulim}
{\bar \nu}_{\pm}&=&\pm\frac{\sqrt{2}}4\left[\frac{\Sigma\left[\Sigma+\sqrt{\Sigma^2+24 k_{\rm (lie)}{}_{\hat \rho}^2\nu_\pm^4}\right]+4 k_{\rm (lie)}{}_{\hat \rho}^2\nu_\pm^4}{k_{\rm (lie)}{}_{\hat \rho}(\partial_{\hat \rho}\ln{k_{\rm (lie)}{}_{\hat \rho}}-k_{\rm (lie)}{}_{\hat \rho})}\right]^{1/2}\nonumber\\
\Sigma&=&2\partial_{\hat \rho}\ln{k_{\rm (lie)}{}_{\hat \rho}}-k_{\rm (lie)}{}_{\hat \rho}(2+\nu_\pm^2)\ ;
\end{eqnarray}
moreover, the timelike condition for $|\nu_p| <1$ is satisfied for all values of $\nu$ outside the same interval.

A linear relation between $\nu$ and $s$ can be obtained in the limit of small $s$:
\begin{equation}
\label{solTexpnu}
\nu= \pm\nu_\pm+{\mathcal N}^{(T)}s+O(s^2)\ , \qquad {\mathcal N}^{(T)}\equiv{\mathcal N}^{(P)}\ .
\end{equation}
From this approximate solution for $\nu$ we also have that 
\begin{equation}
\nu_p^{(\pm)}= \pm\nu_\pm+{\mathcal N}^{(T)}s+O(s^2)\ , 
\end{equation}
and so the total 4-momentum $P$ is given by equation~(\ref{Ptot}) with 
\begin{equation}
\nu_p=\nu+O(s^2)\ .
\end{equation}
The angular velocities $\zeta$, $\zeta_p$ and their reciprocals coincide with the corresponding ones derived in the case of P supplementary conditions (see equations~(\ref{zetaP}) and (\ref{zetapP}) respectively).

\section{Clock effect for spinning test particles}
\label{clocksect}

As we have  seen in all cases examined above, when the circular motion of spinning test particles is considered on particular symmetry hyperplanes 
corresponding to the condition $k_{\rm (lie)}{}_{\hat z}=0$, the orbits are close to a geodesic (as expected) for small values of the spin $s$, with
\beq
\label{clock}
\frac{1}{\zeta_{(SC,\pm,\pm)}}=\pm \frac{1}{\zeta_\pm} \pm |s| {\mathcal J}_{SC}\ ,
\eeq 
where 
\beq
\label{Jsc}
{\mathcal J}_{SC}=-\frac{e^{2\psi}}{\rho\zeta_\pm^2}{\mathcal N}^{(SC)}\ .
\eeq
Equation~(\ref{clock}) identifies these orbits according to the chosen supplementary conditions, the signs in front of $1/\zeta_\pm$ 
corresponding to orbits which co/counter rotate with respect to a pre-assigned sense of variation of the azimuthal angle $\phi$,
 while the signs in front of $s$ refer to a positive or negative spin direction along the $z$-axis; for instance, the quantity 
$\zeta_{(P,+,-)}$ denotes the angular velocity of $U$, 
derived under the choice of Pirani's supplementary conditions and corresponding to a co-rotating orbit $(+)$ with spin-down $(-)$ alignment, etc.
Therefore one can measure the difference in the arrival times due to the spin after one complete revolution with respect to a 
static observer, i.e. what is called gravitomagnetic \lq\lq clock effect''.
This effect has already been studied in Schwarzschild and Kerr spacetimes \cite{bdfg1,bdfg2}. The coordinate time difference is given by: 
\beq\label{deltat}
\Delta t_{(+,+;-,+)}= 2\pi \left(\frac{1}{\zeta_{(SC,+,+)}}+\frac{1}{\zeta_{(SC,-,+)}}\right)=4 \pi |s| {\mathcal J}_{SC}\ ,
\eeq
and analogously for $\Delta t_{(+,-;-,-)}$. This time difference can, in principle, be measured giving some hints for the whole model of spinning test particles in General Relativity. 
In the next section we shall give explicit examples for superposed Weyl fields corresponding to Chazy-Curzon particles and  Schawarzschild black holes. The values of ${\mathcal J}_{SC}$ for the Weyl solutions here examined are explicitly listed in the following section.

\section{Applications}

Our aim now is to apply the theory developed in the previous sections to the static vacuum spacetimes belonging to the Weyl class and 
representing the field of a Chazy-Curzon particle or a Schwarzschild black hole, as well as superpositions of them.
We shall identify the whole class of spatially circular orbits compatible with given values of the spin $s$ and the linear tangential velocity $\nu$. We also find for each of the circular orbits the parameters which allow one to evaluate the clock effect which has a direct physical meaning.
Most of the results, however, will be  discussed with the aid of plots because of the very long formulas involved in the treatment of such solutions.


Let us start by describing solutions belonging to the Weyl class \cite{weyl,exactsols} and representing superpositions of two or more axially symmetric bodies. In general these solutions correspond to configurations which are not gravitationally stable; 
this fact is revealed by the occurrence of gravitationally inert singular
structures (``struts'' and ``membranes'') that keep the bodies apart making the configuration stable (see, e.g. \cite{letelier} and references therein).
In what follows we list the metric coefficients for the examined solutions as well as the relevant quantities to evaluate the clock effect on mirror symmetry hyperplane ($k_{\rm (lie)}{}_{\hat z}=0$), as  pointed out in Section \ref{clocksect}.

\begin{enumerate}

{\bf 
\item[1.] The single Chazy-Curzon particle
}

A single Chazy-Curzon particle is a static axisymmetric solution of Einstein's equations endowed with a naked singularity at the particle position \cite{chazy,curzon,scott}. 
The Curzon metric is generated by the newtonian potential of a spherically symmetric point mass using the Weyl formalism;  the metric coefficients in (\ref{weylmetric}) read
\begin{equation}\fl\quad
\label{ccsolw}
\psi_{\rm C}=-\frac{M_{\rm C}}{R_{\rm C}}\ , \qquad \gamma_{\rm C}=-\frac12\frac{M_{\rm C}^2\rho^2}{R_{\rm C}^4}\ , \qquad R_{\rm C}=\sqrt{\rho^2+z^2}\ . 
\end{equation}

We obtain
\begin{eqnarray}
\nu_\pm&=&\pm\left[\frac{M_{\rm C}}{\rho-M_{\rm C}}\right]^{1/2}\ , \nonumber\\
{\mathcal N}^{(CP)}&=&-\frac{1}{2m}\frac{M_{\rm C}^2}{\rho^3}\frac{\nu_\pm^2}{\gamma_\pm}e^{-\frac12\frac{M_{\rm C}}{\rho^2}(2\rho-M_{\rm C})}\ , \nonumber\\
{\mathcal N}^{(P)}&=&{\mathcal N}^{(CP)}\gamma_\pm\left[\frac{3}{\nu_\pm^2}\frac{\rho}{M_{\rm C}}+2\right]\ , 
\end{eqnarray}
so that 
\begin{eqnarray}
{\mathcal J}_{CP}&=&\frac{1}{2m}\frac{M_{\rm C}^2}{\rho^2}\frac{1}{\gamma_\pm}e^{\frac12\frac{M_{\rm C}}{\rho^2}(2\rho+M_{\rm C})}\ , \nonumber\\
{\mathcal J}_{P}&=&{\mathcal J}_{CP}\,\gamma_\pm\left[\frac{3}{\nu_\pm^2}\frac{\rho}{M_{\rm C}}+2\right]\ .
\end{eqnarray}

{\bf 
\item[2.] Superposition of two Chazy-Curzon particles
}

The solution corresponding to the superposition of two Chazy-Curzon particles with masses $M_{\rm C}$ and $m_{\rm C_b}$ and positions $z=0$ and $z=b$ on the $z$-axis respectively is given by metric (\ref{weylmetric}) with functions
\begin{equation}
\label{psigammaCCb}
\psi=\psi_{\rm C}+\psi_{\rm C_b}\ , \qquad \gamma=\gamma_{\rm C}+\gamma_{\rm C_b}+\gamma_{\rm CC_b}\ ,
\end{equation}
where $\psi_{\rm C}$, $\gamma_{\rm C}$ are defined by equations~(\ref{ccsolw}), while
\begin{equation}\fl\quad
\label{ccbsolw}
\psi_{\rm C_b}=-\frac{m_{\rm C_b}}{R_{\rm C_b}}\ , \qquad \gamma_{\rm C_b}=-\frac12\frac{m_{\rm C_b}^2\rho^2}{R_{\rm C_b}^4}\ , \qquad R_{\rm C_b}=\sqrt{\rho^2+(z-b)^2}\ 
\end{equation}
and $\gamma_{\rm CC_b}$ can be obtained by solving Einstein's equations (\ref{einsteqs}):
\begin{equation}
\gamma_{\rm CC_b}=2\frac{m_{\rm C_b}M_{\rm C}}{b^2}\frac{\rho^2+z(z-b)}{R_{\rm C_b}R_{\rm C}}+C\ .
\end{equation}
The value of the arbitrary constant $C$ can be determined by imposing the regularity condition
\begin{equation}
\label{regcond}
\lim_{\rho\rightarrow0}\gamma=0\ ;
\end{equation}
however, it cannot be uniquely chosen in order to make the function $\gamma_{\rm CC_b}$  vanishing on the whole $z$-axis: a $\gamma_{\rm CC_b}\not= 0$ gives rise to a conical singularity (see, e.g. \cite{sokolov,israel}), corresponding to a strut in compression, which holds the two particles apart. 
The choice $C=2m_{\rm C_b}M_{\rm C}/b^2$ makes $\gamma_{\rm CC_b}=0$ only on the segment $0<z<b$ of the $z$-axis between the sources. 
In the following we use $C=-2m_{\rm C_b}M_{\rm C}/b^2$, that makes $\gamma_{\rm CC_b}=0$ on the portion of the axis with $z<0$ and $z>b$.

We obtain
\begin{eqnarray}\fl\quad
\nu_\pm&=&\pm 4\rho\left[\frac{M_{\rm C}}{{\mathcal R}^2({\mathcal R}-4M_{\rm C})+4M_{\rm C}b^2}\right]^{1/2}\ , \nonumber\\
\fl\quad
{\mathcal N}^{(CP)}&=&-\frac{1}{2m}\frac{\nu_\pm^2}{\rho}\left[2-\frac{{\mathcal R}^3\nu_\pm^2}{4M_{\rm C}({\mathcal R}^2-b^2)}\right]^{1/2}[{\mathcal R}^4(16M_{\rm C}^2-3b^2)\nonumber\\
\fl\quad
&&-16M_{\rm C}^2b^2(2{\mathcal R}^2-b^2)]\frac{e^{-\frac{4M_{\rm C}}{{\mathcal R}^4}[{\mathcal R}^2({\mathcal R}-2M_{\rm C})+M_{\rm C}b^2]}}{{\mathcal R}^6}\ , \nonumber\\
\fl\quad
{\mathcal N}^{(P)}&=&
{\mathcal N}^{(CP)}\frac{{\mathcal R}^4(3{\mathcal R}^2+32M_{\rm C}^2-6b^2)-12M_{\rm C}({\mathcal R}^2-b^2)({\mathcal R}^3+4M_{\rm C}b^2)}{{\mathcal R}^4(16M_{\rm C}^2-3b^2)-16M_{\rm C}^2b^2(2{\mathcal R}^2-b^2)}\cdot \nonumber\\
\fl\quad
&&\cdot\left[2-\frac{{\mathcal R}^3\nu_\pm^2}{4M_{\rm C}({\mathcal R}^2-b^2)}\right]^{-1/2}\ , 
\end{eqnarray}
where ${\mathcal R}=\sqrt{4\rho^2+b^2}$, so that 
\begin{eqnarray}\fl\quad
{\mathcal J}_{CP}&=&\frac{1}{2m}\left[2-\frac{{\mathcal R}^3\nu_\pm^2}{4M_{\rm C}({\mathcal R}^2-b^2)}\right]^{1/2}[{\mathcal R}^4(16M_{\rm C}^2-3b^2)-16M_{\rm C}^2b^2(2{\mathcal R}^2-b^2)]\cdot \nonumber\\
\fl\quad
&&\cdot\frac{e^{\frac{4M_{\rm C}}{{\mathcal R}^4}[{\mathcal R}^2({\mathcal R}+2M_{\rm C})-M_{\rm C}b^2]}}{{\mathcal R}^6}\ , \nonumber\\
\fl\quad
{\mathcal J}_{P}&=&{\mathcal J}_{CP}\frac{{\mathcal R}^4(3{\mathcal R}^2+32M_{\rm C}^2-6b^2)-12M_{\rm C}({\mathcal R}^2-b^2)({\mathcal R}^3+4M_{\rm C}b^2)}{{\mathcal R}^4(16M_{\rm C}^2-3b^2)-16M_{\rm C}^2b^2(2{\mathcal R}^2-b^2)}\cdot \nonumber\\
\fl\quad
&&\cdot\left[2-\frac{{\mathcal R}^3\nu_\pm^2}{4M_{\rm C}({\mathcal R}^2-b^2)}\right]^{-1/2}\ . 
\end{eqnarray}

{\bf 
\item[3.] The single Schwarzschild black hole
}

The Schwarzschild black hole solution is generated by the newtonian potential of a line source (a homogeneous rod) of mass $M_{\rm S}$ and lenght $2L$ (with the further position $L=M_{\rm S}$) lying on the axis and placed symmetrically with respect to the origin:
\begin{eqnarray}\fl\quad
\label{Ssolw}
\psi_{\rm S}&=&\frac12\ln{\left[\frac{R_1^{+}+R_1^{-}-2M_{\rm S}}{R_1^{+}+R_1^{-}+2M_{\rm S}}\right]}\ , \qquad
\gamma_{\rm S}=\frac12\ln{\left[\frac{(R_1^{+}+R_1^{-})^2-4M_{\rm S}^2}{4R_1^{+}R_1^{-}}\right]}\ , 
\end{eqnarray}
where
\beq
R_1^{\pm}=\sqrt{\rho^2+(z\pm M_{\rm S})^2}\ .
\eeq
The more familiar form of the Schwarzschild solution in Schwarzschild coordinates ($t,r, \theta, \phi$) is recovered by performing the coordinate transformation
\beq
\rho= \sqrt{r^2-2M_{\rm S}r}\sin \theta, \qquad z=(r-M_{\rm S})\cos \theta .
\eeq

We obtain
\begin{eqnarray}
\nu_\pm&=&\pm\left[\frac{M_{\rm S}}{\sqrt{\rho^2+M_{\rm S}^2}-M_{\rm S}}\right]^{1/2}\ , \nonumber\\
{\mathcal N}^{(CP)}&=&0\ , \qquad
{\mathcal N}^{(P)}= -\frac{3}{2}\frac{M_{\rm S}^2}{m \nu_\pm^2}\frac1{\rho^3}\ ,
%
\end{eqnarray}
so that 
\begin{eqnarray}
{\mathcal J}_{CP}=0\ , \qquad {\mathcal J}_{P}&=&\frac{3}{2}\frac{1}{m}\ .
\end{eqnarray}
Note that in this case no clock-effect is found if the CP supplementary conditions are imposed.

{\bf 
\item[4.] Superposition of two Schwarzschild black holes
}

The solution corresponding to a linear superposition of two Schwarzschild black holes with masses $M_{\rm S}$ and $m_{\rm S_b}$ and positions $z=0$ and $z=b$ on the $z$-axis respectively is given by metric (\ref{weylmetric}) with functions
\begin{eqnarray}
\label{psigammaSSb}
\psi=\psi_{\rm S}+\psi_{\rm S_b}\ , \qquad \gamma=\gamma_{\rm S}+\gamma_{\rm S_b}+\gamma_{\rm SS_b}\ ,
\end{eqnarray}
where $\psi_{\rm S}$, $\gamma_{\rm S}$ are defined by equations~(\ref{Ssolw}), while
\begin{eqnarray}\fl\,
\label{SSbsol}
\psi_{\rm S_b}&=&\frac12\ln{\left[\frac{R_2^{+}+R_2^{-}-2m_{\rm S_b}}{R_2^{+}+R_2^{-}+2m_{\rm S_b}}\right]}\ , \qquad
\gamma_{\rm S_b}=\frac12\ln{\left[\frac{(R_2^{+}+R_2^{-})^2-4m_{\rm S_b}^2}{4R_2^{+}R_2^{-}}\right]}\ , \nonumber\\
\fl\,
\gamma_{\rm SS_b}&=&\frac12\ln{\left[\frac{E_{(1^{+},2^{-})}E_{(1^{-},2^{+})}}{E_{(1^{+},2^{+})}E_{(1^{-},2^{-})}}\right]}+C \ , \quad
E_{(1^{\pm},2^{\pm})}=\rho^2+R_1^{\pm}R_2^{\pm}+Z_1^{\pm}Z_2^{\pm}\ ,
\end{eqnarray}
 where
\begin{eqnarray}
R_1^{\pm}&=&\sqrt{\rho^2+(Z_1^{\pm})^2}\ , \qquad
R_2^{\pm}=\sqrt{\rho^2+(Z_2^{\pm})^2}\ , \nonumber\\
Z_1^{\pm}&=&z\pm M_{\rm S}\ , \qquad
Z_2^{\pm}=z-(b\mp m_{\rm S_b})\ .
\end{eqnarray}
The function $\gamma_{\rm SS_b}$ is obtained by solving Einstein's equations (\ref{einsteqs}).
The value of arbitrary constant $C$ can be determined by imposing the regularity condition (\ref{regcond}); we make the choice $C=0$, so that the function $\gamma_{\rm SS_b}$ vanishes on the portions of the $z$-axis outside the sources (that is, for $z>b+m_{\rm S_b}$ and $z<-M_{\rm S}$).

We obtain
\begin{eqnarray}\fl\quad
\nu_\pm&=&\pm 4\rho\left[\frac{M_{\rm S}({\mathcal R}_{+}+{\mathcal R}_{-})}{{\mathcal B}_1}\right]^{1/2}\ , \nonumber\\
\fl\quad
{\mathcal N}^{(CP)}&=&-768\rho\frac{M_{\rm S}}{m}\frac{b^2-4M_{\rm S}^2}{({\mathcal R}_{+}+{\mathcal R}_{-}+4M_{\rm S})^2[(4\rho^2+b^2+{\mathcal R}_{+}{\mathcal R}_{-})^2-16M_{\rm S}^4]}\frac{{\mathcal B}_3}{{\mathcal B}_1^{3/2}{\mathcal B}_2^{1/2}}\ , \nonumber\\
\fl\quad
{\mathcal N}^{(P)}&=&\frac{{\mathcal N}^{(CP)}}{2}\frac{(4\rho^2+b^2+{\mathcal R}_{+}{\mathcal R}_{-})^2-16M_{\rm S}^4}{b^2-4M_{\rm S}^2}\frac{{\mathcal R}_{+}^2{\mathcal R}_{-}^2}{(1+\nu_\pm^2)^2}\frac{{\mathcal B}_4}{{\mathcal B}_3}\frac{{\mathcal B}_2^{1/2}}{{\mathcal B}_1^{3/2}}\ , 
\end{eqnarray}
where the quantities ${\mathcal R}_{\pm}$ and ${\mathcal B}_i$ ($i=1...4$) are defined by
\begin{eqnarray}\fl\quad
{\mathcal R}_{\pm}&=&[4\rho^2+(b\pm2M_{\rm S})^2]^{1/2}\ , \nonumber\\
\fl\quad
{\mathcal B}_1&=&{\mathcal R}_{+}{\mathcal R}_{-}[{\mathcal R}_{+}{\mathcal R}_{-}+4\rho^2+b^2-4M_{\rm S}^2]-16M_{\rm S}\rho^2({\mathcal R}_{+}+{\mathcal R}_{-})\ , \nonumber\\
\fl\quad
{\mathcal B}_2&=&{\mathcal B}_1-16M_{\rm S}\rho^2({\mathcal R}_{+}+{\mathcal R}_{-})\ , \nonumber\\
\fl\quad
{\mathcal B}_3&=&-16M_{\rm S}\rho^2[{\mathcal R}_{+}^2{\mathcal R}_{-}^2+4M_{\rm S}^2b^2]({\mathcal R}_{+}+{\mathcal R}_{-})^2
+{\mathcal R}_{+}{\mathcal R}_{-}\big\{
-256 M_{\rm S}^3\rho^2b^2\nonumber\\
\fl\quad
&&+({\mathcal R}_{+}+{\mathcal R}_{-})[{\mathcal R}_{+}^3{\mathcal R}_{-}^3-32 M_{\rm S}^2\rho^2(2\rho^2+4M_{\rm S}^2)+4M_{\rm S}^2(b^4+4M_{\rm S}^2b^2\nonumber\\
\fl\quad
&&-16M_{\rm S}^4)]+2M_{\rm S}b({\mathcal R}_{+}-{\mathcal R}_{-})[{\mathcal R}_{+}^2{\mathcal R}_{-}^2+16 M_{\rm S}^2\rho^2-8M_{\rm S}^2(b^2-2M_{\rm S}^2)]
\big\}\ , \nonumber\\
\fl\quad
{\mathcal B}_4&=&8M_{\rm S}\rho^2[({\mathcal R}_{+}+{\mathcal R}_{-})^2{\mathcal R}_{+}{\mathcal R}_{-}+16M_{\rm S}^2b^2]+({\mathcal R}_{+}+{\mathcal R}_{-})[-{\mathcal R}_{+}^3{\mathcal R}_{-}^3\nonumber\\
\fl\quad
&&+32(b^2-2M_{\rm S}^2)\rho^4-16b^2\rho^2(b^2-M_{\rm S}^2)+2(b^2-4M_{\rm S}^2)(b^4-8M_{\rm S}^4)]\nonumber\\
\fl\quad
&&-2M_{\rm S}b({\mathcal R}_{+}-{\mathcal R}_{-})[{\mathcal R}_{+}^2{\mathcal R}_{-}^2+2(b^2-2M_{\rm S}^2)(4\rho^2+b^2-4M_{\rm S}^2]\ .
\end{eqnarray}
Thus we get
\begin{eqnarray}\fl\quad
{\mathcal J}_{CP}&=&\frac{48}{m}\frac{b^2-4M_{\rm S}^2}{({\mathcal R}_{+}+{\mathcal R}_{-})^2-16M_{\rm S}^2}\frac1{(4\rho^2+b^2+{\mathcal R}_{+}{\mathcal R}_{-})^2-16M_{\rm S}^4}\cdot\nonumber\\
\fl\quad
&&\cdot\frac{{\mathcal B}_3}{({\mathcal B}_1{\mathcal B}_2)^{1/2}}\frac1{{\mathcal R}_{+}+{\mathcal R}_{-}}\ , \nonumber\\
\fl\quad
{\mathcal J}_{P}&=&\frac{{\mathcal J}_{CP}}{2}\frac{(4\rho^2+b^2+{\mathcal R}_{+}{\mathcal R}_{-})^2-16M_{\rm S}^4}{b^2-4M_{\rm S}^2}\frac{{\mathcal R}_{+}^2{\mathcal R}_{-}^2}{(1+\nu_\pm^2)^2}\frac{{\mathcal B}_4}{{\mathcal B}_3}\frac{{\mathcal B}_2^{1/2}}{{\mathcal B}_1^{3/2}}\ .
\end{eqnarray}

\end{enumerate}

\begin{figure}
\typeout{*** EPS figure 1}
\begin{center}
$
\begin{array}{ccc}
\includegraphics[scale=0.29]{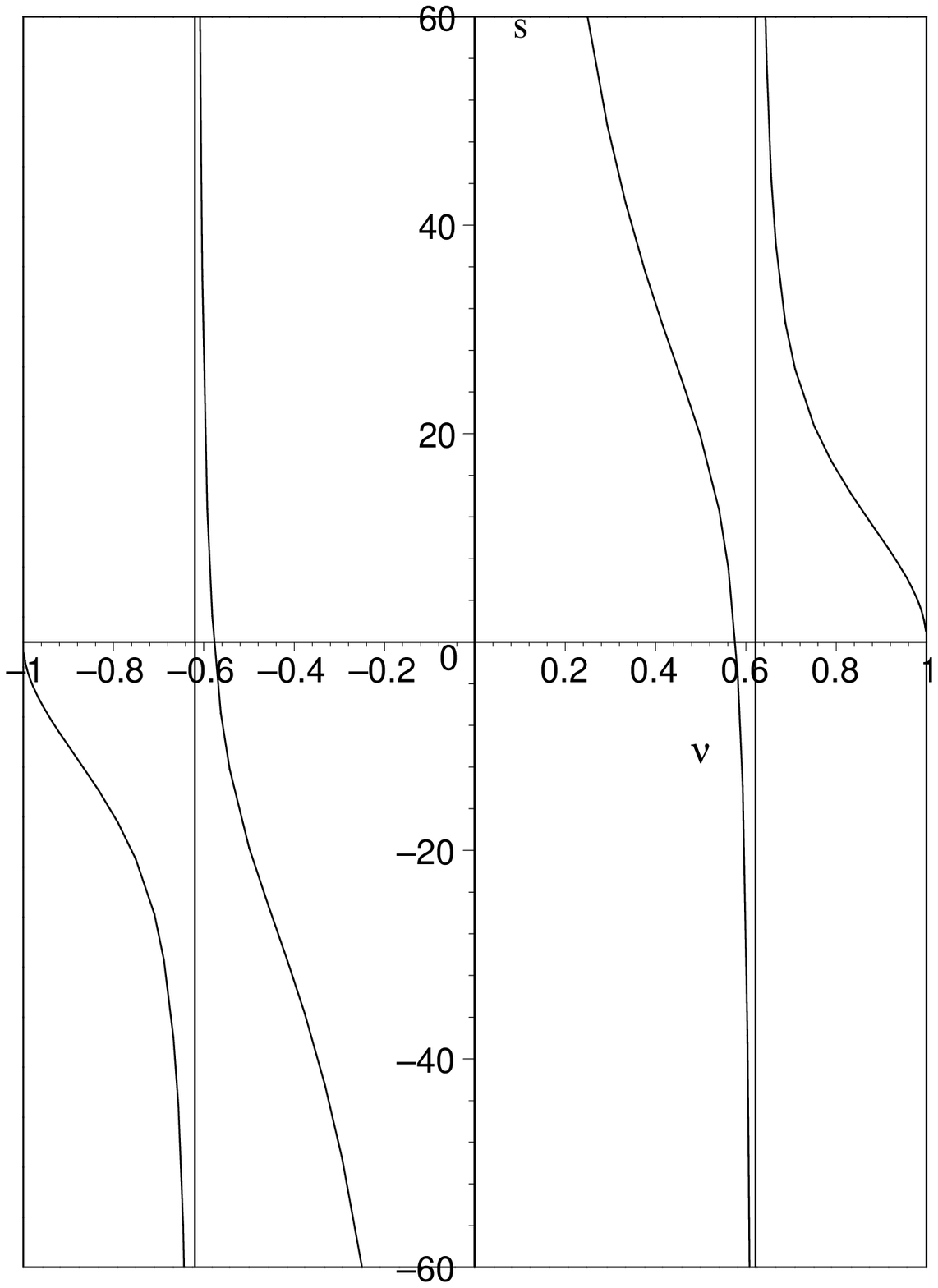}&\includegraphics[scale=0.29]{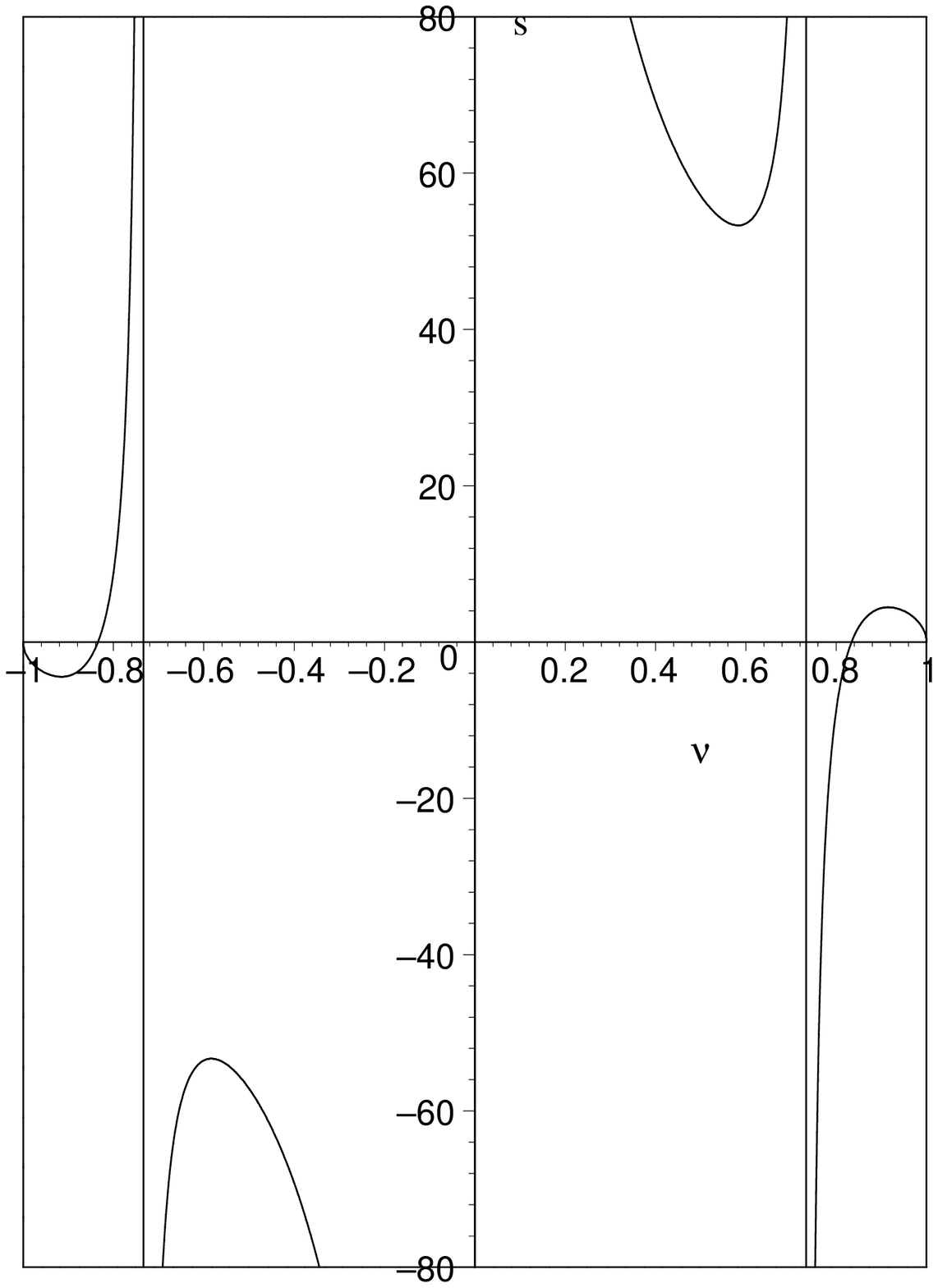}& \includegraphics[scale=0.29]{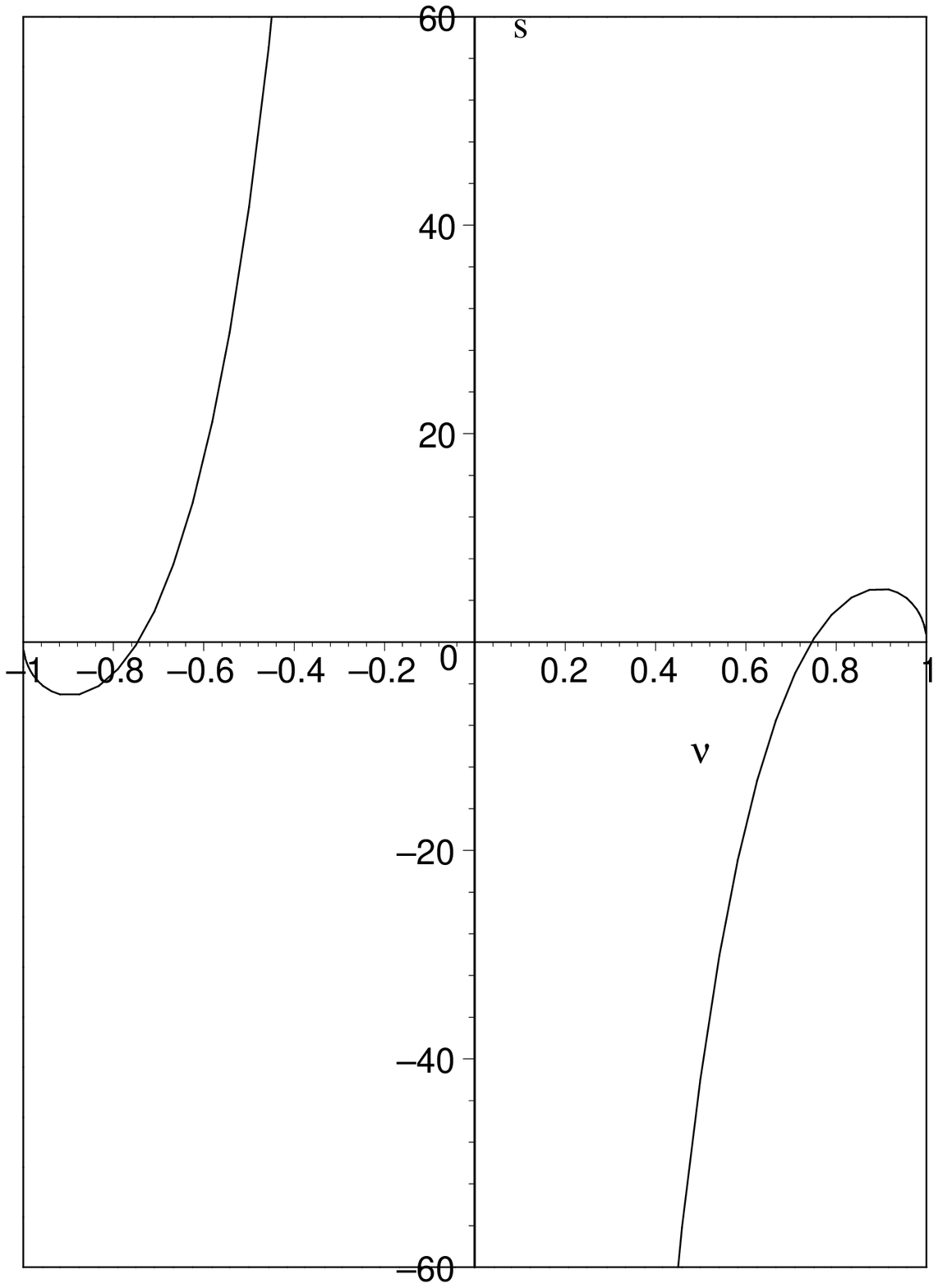}\\[.4cm]
\quad\mbox{(a)}\quad &\quad \mbox{(b)}\quad & \quad\mbox{(c)}
\end{array}
$
\end{center}
\caption{
In the case of CP supplementary conditions, the spin parameter ${\hat s}$ is plotted as a function of the linear velocity $\nu$ in the case $k_{\rm (lie)}{}_{\hat z}=0$ at a fixed value of the radial distance $\rho=4$, for the single Chazy-Curzon particle ($M_{\rm C}=1$), 
two Chazy-Curzon particles ($M_{\rm C}=1=m_{\rm C_b}$, $b=3$) and two Schwarzschild black holes ($M_{\rm S}=1=m_{\rm S_b}$, $b=4$) respectively.
The values of the geodesic linear velocity $\nu_\pm$ corresponding to the given choice of the parameters are $\nu_{\pm}\approx\pm0.577$ (case (a)), $\nu_{\pm}\approx\pm0.834$ (case (b)) and $\nu_{\pm}\approx\pm0.746$ (case (c)) respectively.}
\label{fig:1}
\end{figure}

\begin{figure} 
\typeout{*** EPS figure 2}
\begin{center}
$
\begin{array}{cc}
\includegraphics[scale=0.32]{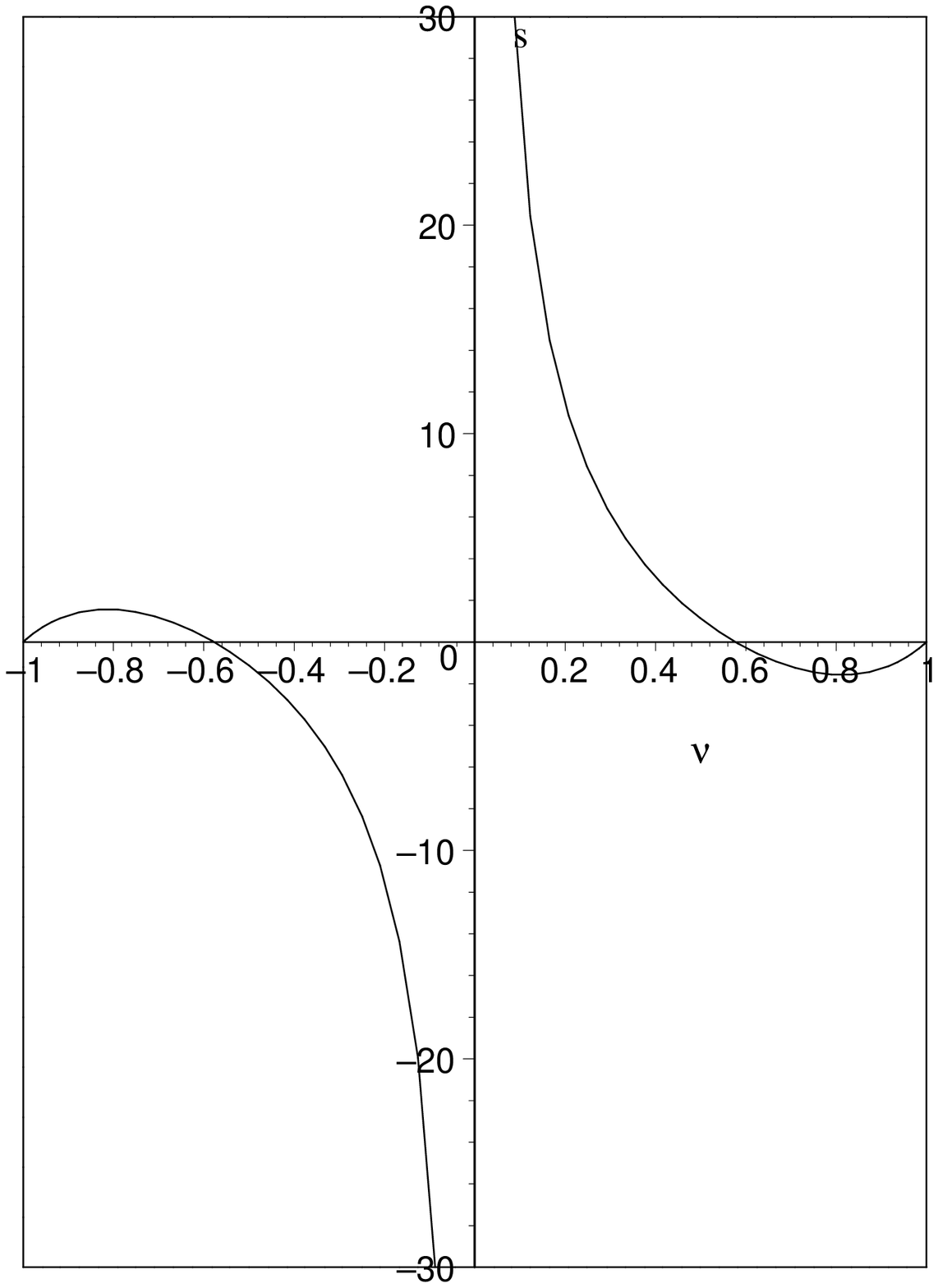}&\quad
\includegraphics[scale=0.32]{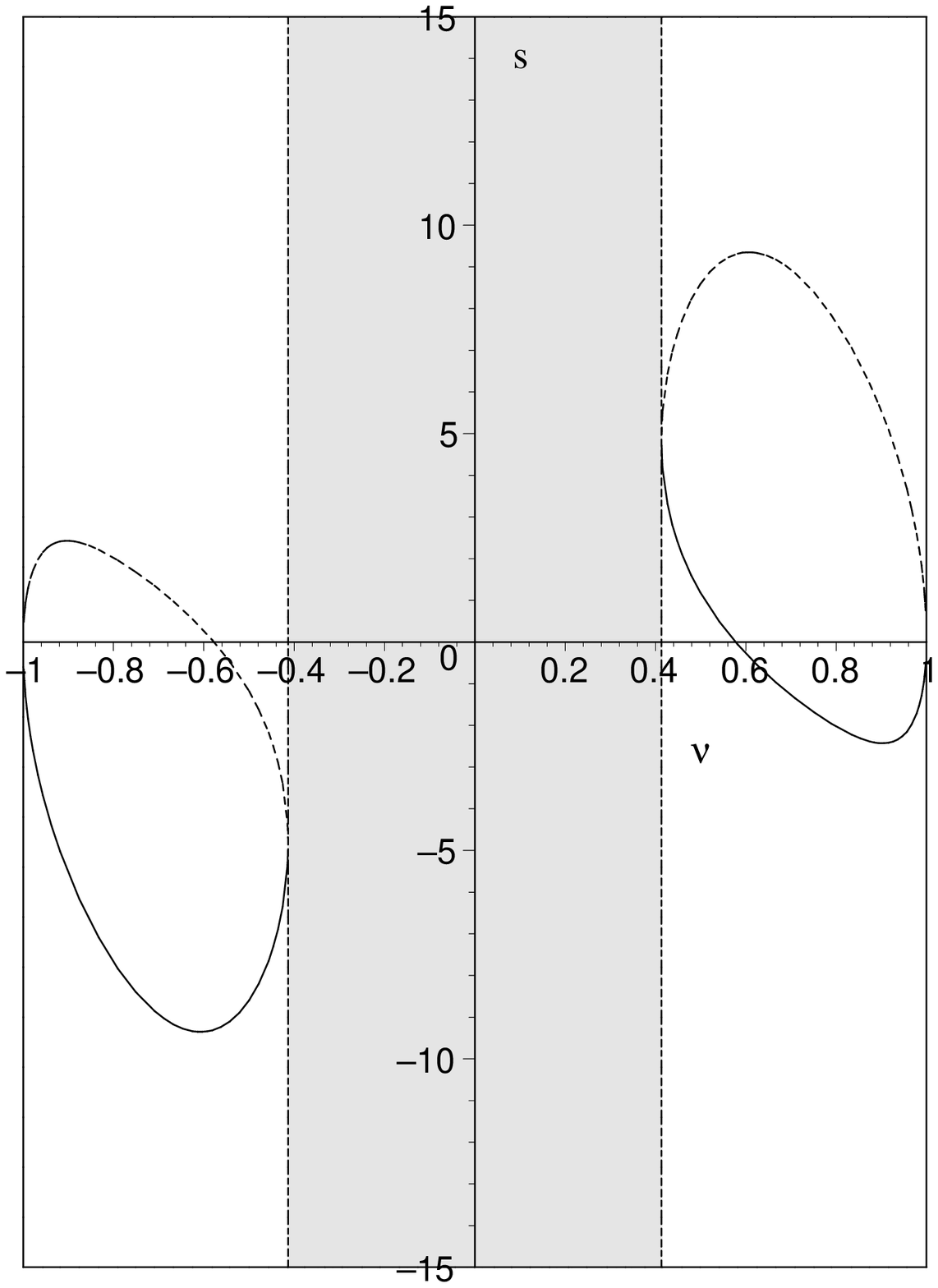}\\[.4cm]
\quad\mbox{(a)}\quad &\quad \mbox{(b)}
\end{array}
$
\end{center}
\caption{
In the case of P and T supplementary conditions (see figures (a) and (b), respectively), the spin parameter ${\hat s}$ is plotted as a function of the linear velocity $\nu$ in the case $k_{\rm (lie)}{}_{\hat z}=0$ at a fixed value of the radial distance $\rho=4$, for the single Chazy-Curzon particle ($M_{\rm C}=1$, and so $\nu_{\pm}\approx\pm0.577$).
The shaded region in the T case contains the forbidden values of $\nu$ (the limiting values are given by ${\bar \nu}_{\pm}\approx\pm0.413$). 
We avoid to show the behaviours corresponding to the other solutions, since they are qualitatively the same. 
We remark that our treatment loses its validity for high values of the spin parameter; hence, the plots of figure (a)  and (b) should be truncated within the range $\hat s \in [-\epsilon, \epsilon]$,  $\epsilon \ll 1$ a dimensionless parameter denoting somehow the physical region.}
\label{fig:2}
\end{figure}

Figures \ref{fig:1}  and  \ref{fig:2}  
show the behaviour of the spin parameter ${\hat s}$ as a function of the linear velocity $\nu$ in the case $k_{\rm (lie)}{}_{\hat z}=0$ for a fixed value of the radial distance $\rho$ and
for each choice of supplementary conditions. 
The spin parameter ${\hat s}=\pm|{\hat s}|=\pm|s|/(m\mu)$ is defined as the signed magnitude of the spin per unit (bare) mass $m$ of the test particle and $\mu=M_{\rm C}$ or $\mu=M_{\rm S}$ of the Chazy-Curzon particle(s)
or Schwarzschild black hole(s).
The symmetry hyperplanes lie at $z=0$ in the case of one-body solutions, and at $z=b/2$ for equal masses $M_1=M_2$ in the case of two-body solutions, with the bodies located at $z=0$ and $z=b$ on the $z$-axis.
As we can see in Figure 1 the CP supplementary
condition appear inadequate since they lead to unphysical situations.
In all cases considered, in fact, spinning particles
at rest ($\nu=0$) require an infinite spin; moreover circular orbits with
an infinite
spin are also found close to a geodesic ($\hat s=0$)
in both the single and two Chazy-Curzon particles solutions. 
The case of a single Schwarzschild black hole is somehow particular, when CP supplementary conditions are imposed: in fact, as widely discussed in \cite{bdfg1}, the only physical solution corresponds to $\nu=\nu_\pm$ and $\hat s$ arbitrary.

Despite the formal complexity there is no significant difference among the  solutions we have considered in the behaviour of the spin as a function of the speed in either P or T supplementary conditions, in contrast with CP case; thus, it is enough to show these behaviours referring to the single Chazy-Curzon solution only (see Figure 2).
As we can see, only
the T supplementary conditions provide physically significant ($\hat s-\nu$) plots
for any value of the spin $\hat s$. In this case, in fact, the spin $\hat s$ is always bounded; however,  the request of smallness for its magnitude can also give restrictions here, in order to model realistic situations.

\begin{figure} 
\typeout{*** EPS figure 3}
\begin{center}
$
\begin{array}{cc}
\includegraphics[scale=0.35]{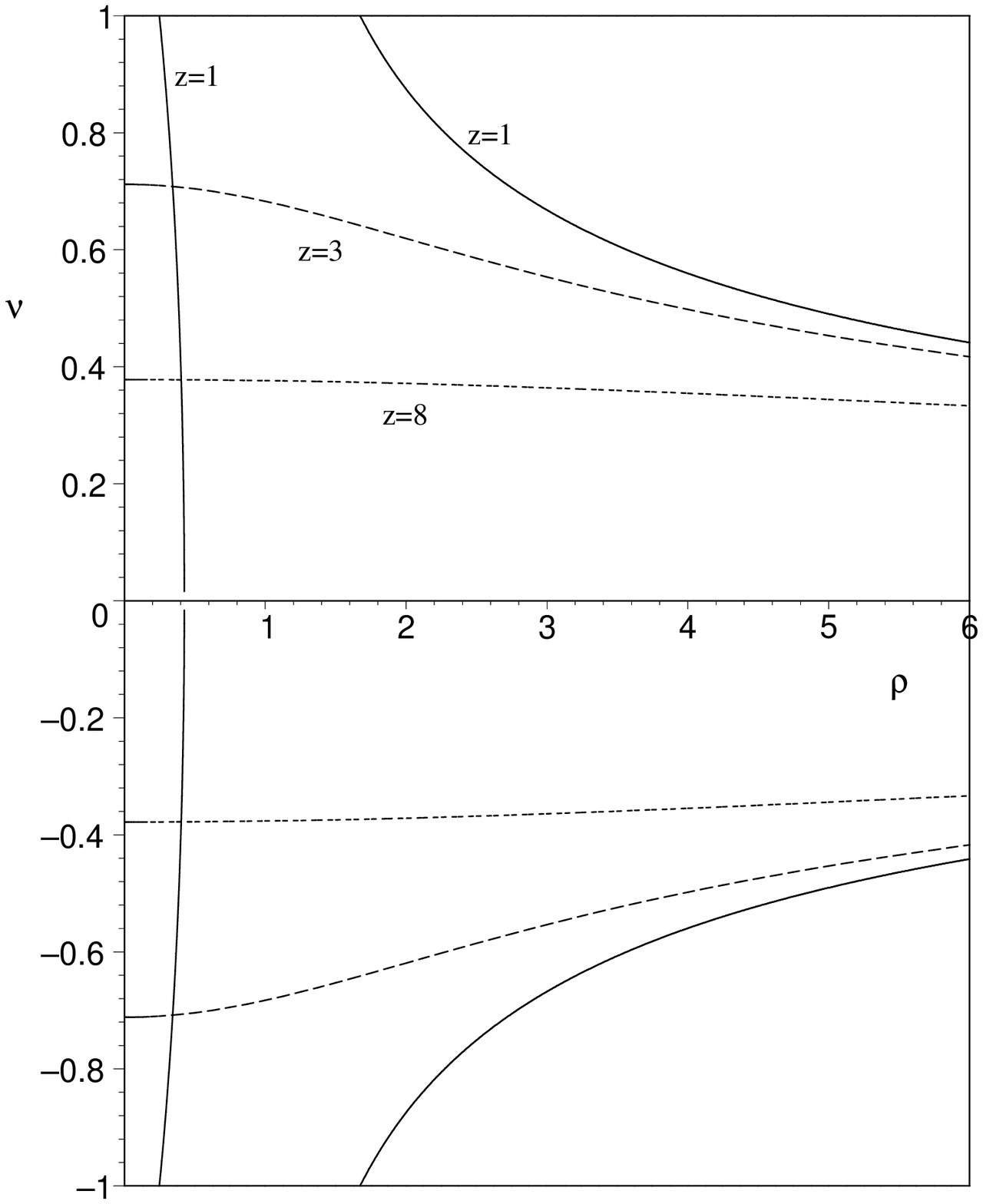}&\quad
\includegraphics[scale=0.35]{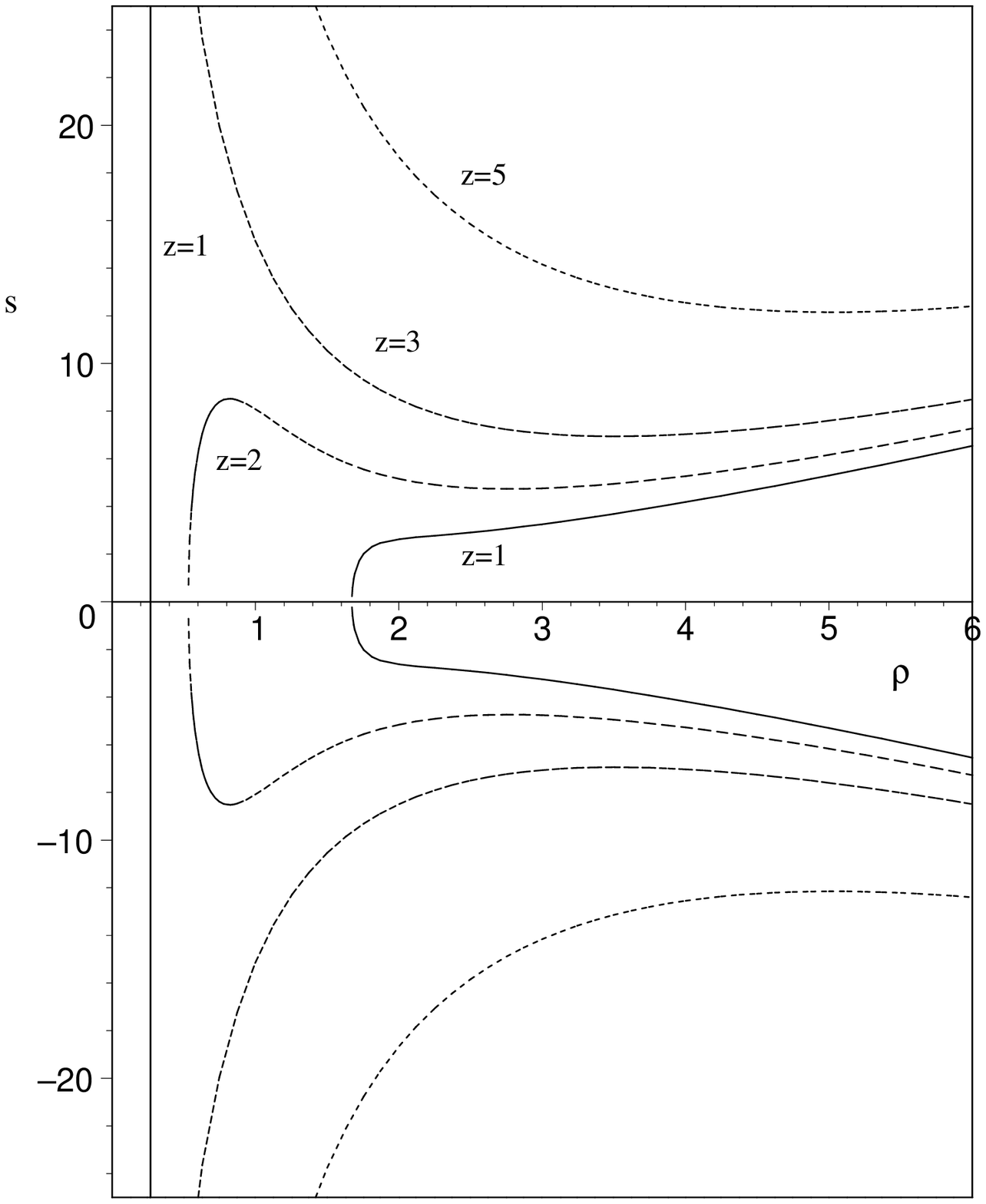}\\[.4cm]
\quad\mbox{(a)}\quad &\quad \mbox{(b)}\\[.6cm]
\includegraphics[scale=0.35]{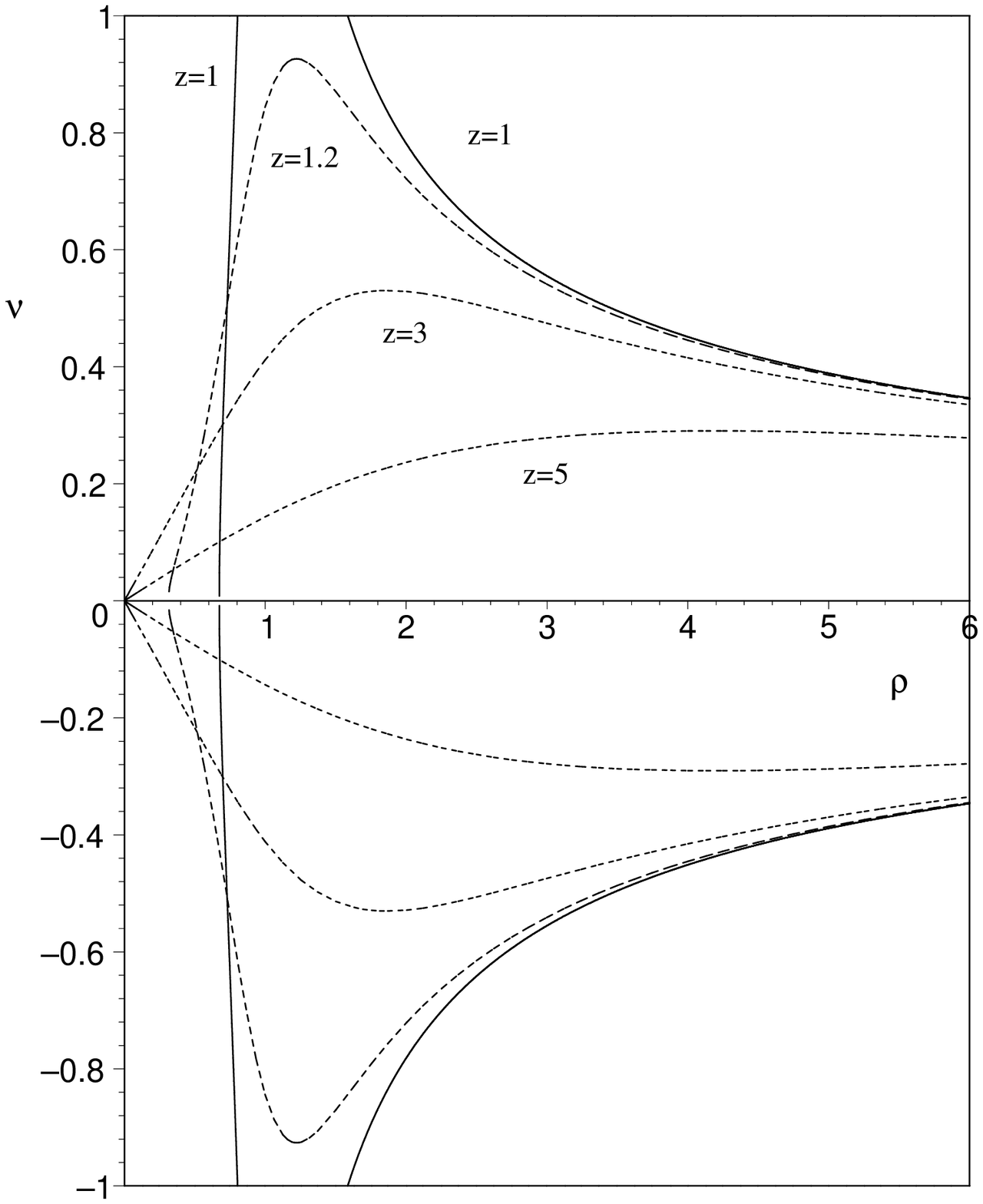}&\quad
\includegraphics[scale=0.35]{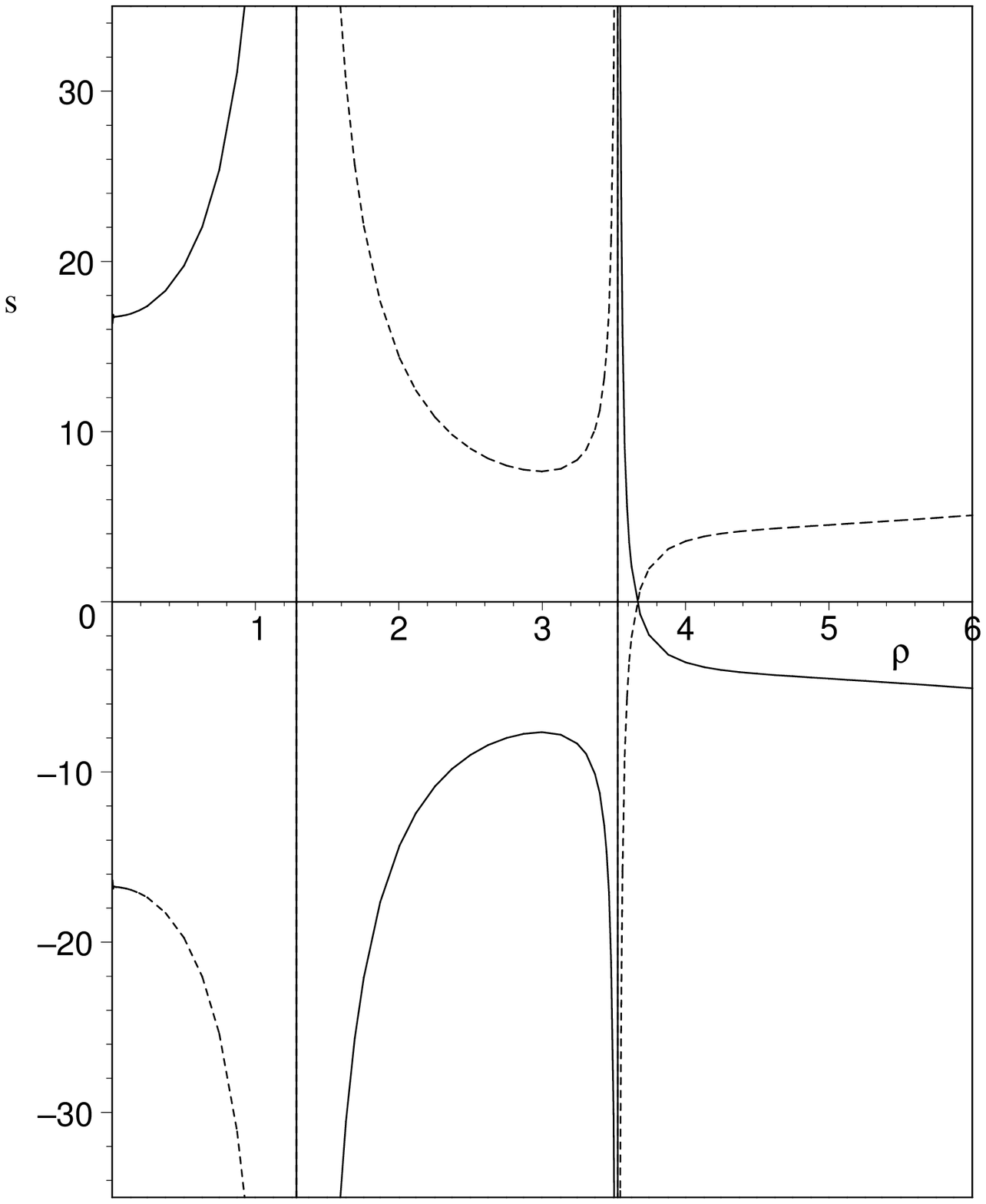}\\[.4cm]
\quad\mbox{(c)}\quad &\quad \mbox{(d)}
\end{array}
$\\
\end{center}
\caption{
In the case of the single Chazy-Curzon particle, the linear velocity ${}^s\nu_{\pm}$ for co/counter-rotating circular orbits and the corresponding spin parameter ${\hat s}$ are evaluated on different planes $z=const$, and plotted in figures (a), (c) and figures (b), (d) respectively as functions of $\rho$ (and $M_{\rm C}=1$),
for both CP (see figures (a), (b)) and P (see figures (c), (d)) supplementary conditions.
A choice of values of $z$ different for each plot has been made for the sake of clarity.
Solid, dotted and dashdotted lines correspond to $z=1,3,8$ respectively in  figure (a); solid, dotted, dashed and dashdotted lines refer to the choices $z=1,2,3,5$ in figure (b) and $z=1,1.2,2,5$ in figure(c); 
in figure (d) we plot only the case $z=3$ as an example, with solid and dotted lines referring to co/counter-rotating orbits respectively.
}
\label{fig:3}
\end{figure}

\begin{figure} 
\typeout{*** EPS figure 4}
\begin{center}
$
\begin{array}{cc}
\includegraphics[scale=0.35]{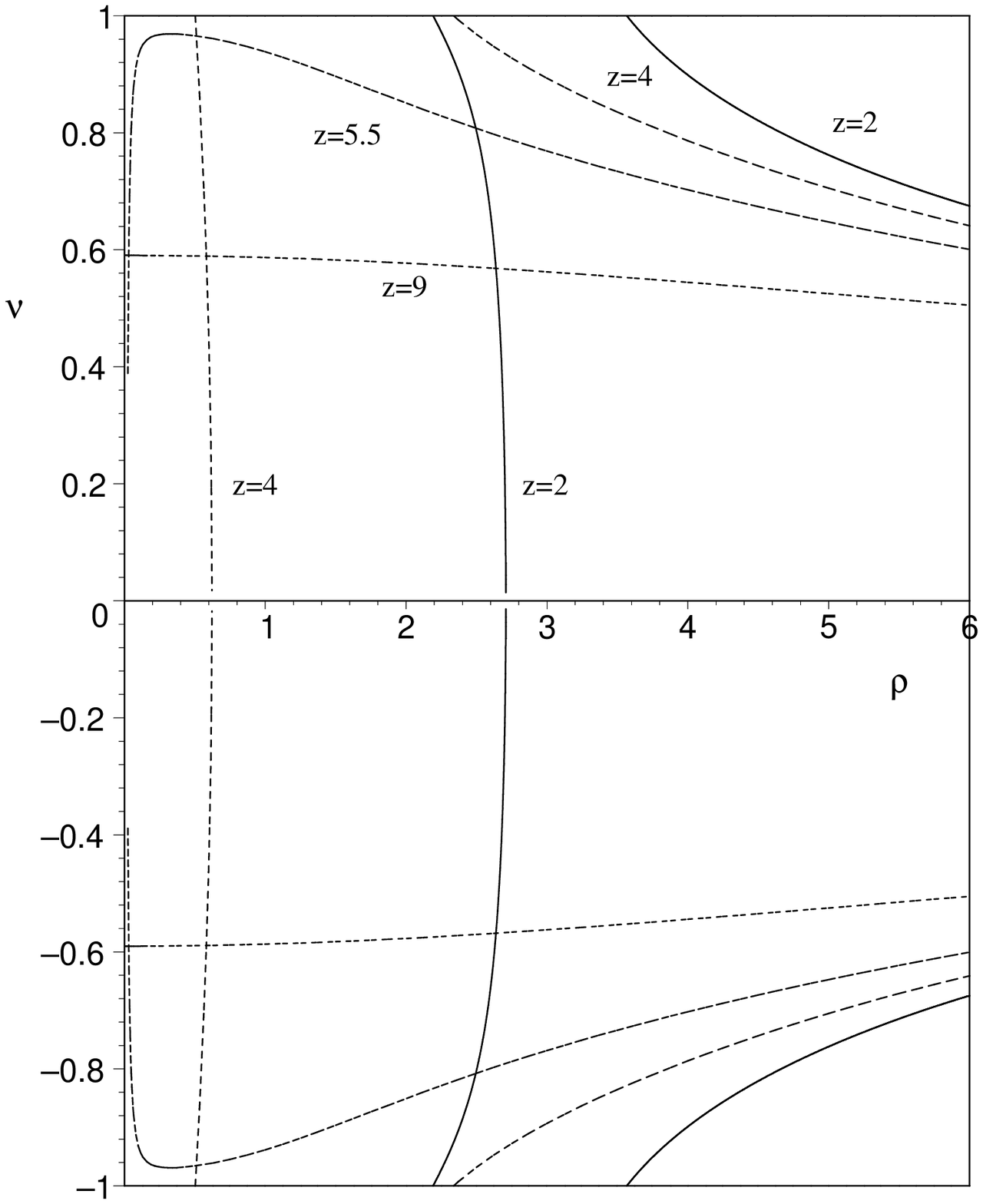}&\quad
\includegraphics[scale=0.35]{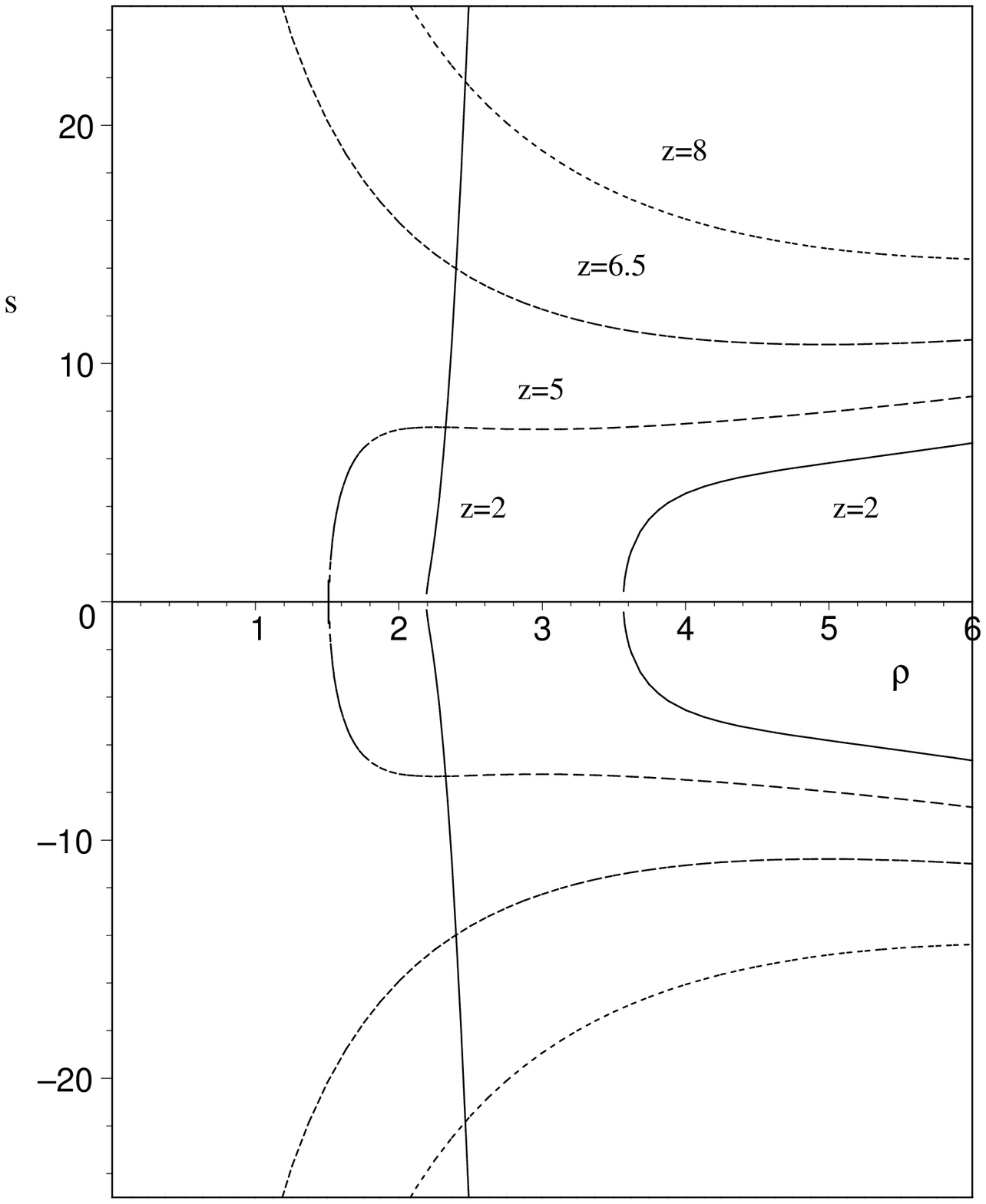}\\[.4cm]
\quad\mbox{(a)}\quad &\quad \mbox{(b)}\\[.6cm]
\includegraphics[scale=0.35]{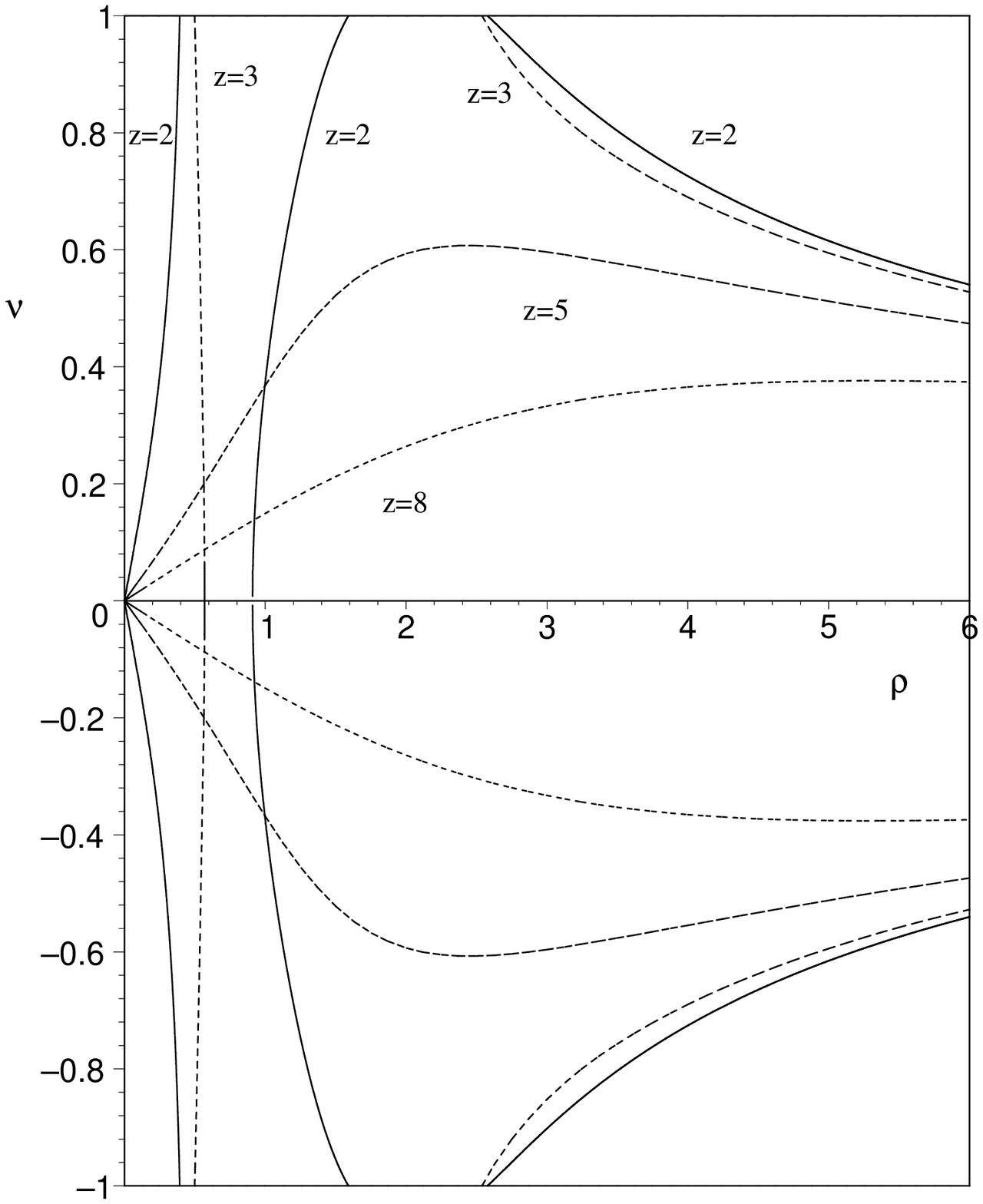}&\quad
\includegraphics[scale=0.35]{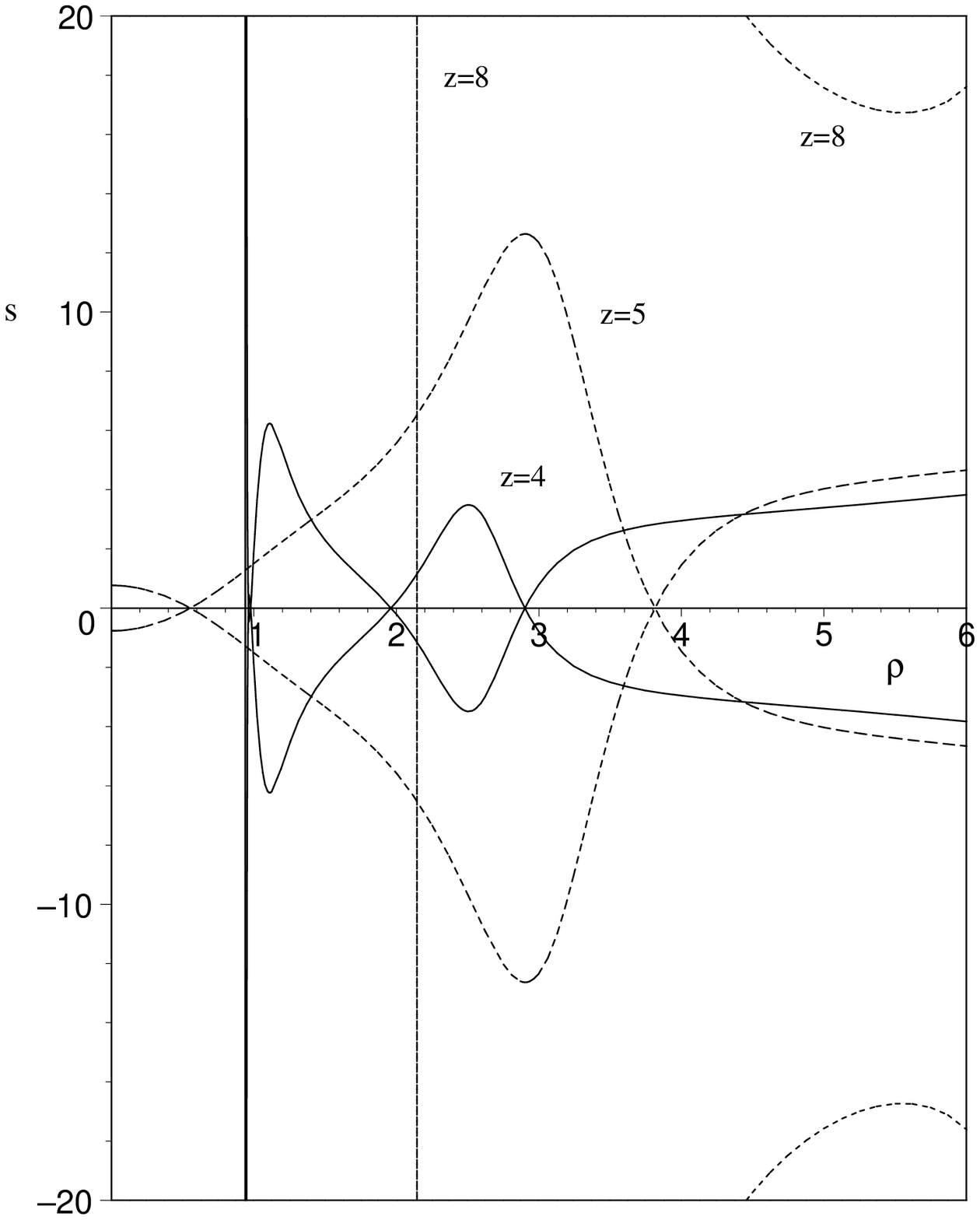}\\[.4cm]
\quad\mbox{(c)}\quad &\quad \mbox{(d)}
\end{array}
$\\
\end{center}
\caption{
In the case of two Chazy-Curzon particles, the linear velocity ${}^s\nu_{\pm}$ for co/counter-rotating circular orbits and the corresponding spin parameter ${\hat s}$ are evaluated on different planes $z=const$, and plotted in figures (a), (c) and figures (b), (d) respectively as functions of $\rho$ (and $M_{\rm C}=1=m_{\rm C_b}$, $b=3$), 
for both CP (see figures (a), (b)) and P (see figures (c), (d)) supplementary conditions.
Solid, dotted, dashed and dashdotted lines correspond to $z=2,4,5.5,9$ respectively in  figure  (a), $z=2,5,6.5,8$ in figure (b) and $z=2,3,5,8$ in figure (c); solid, dotted and dashdotted lines refer to the choice $z=4,5,8$ in figure (d).
}
\label{fig:4}
\end{figure}

\begin{figure} 
\typeout{*** EPS figure 5}
\begin{center}
$
\begin{array}{cc}
\includegraphics[scale=0.35]{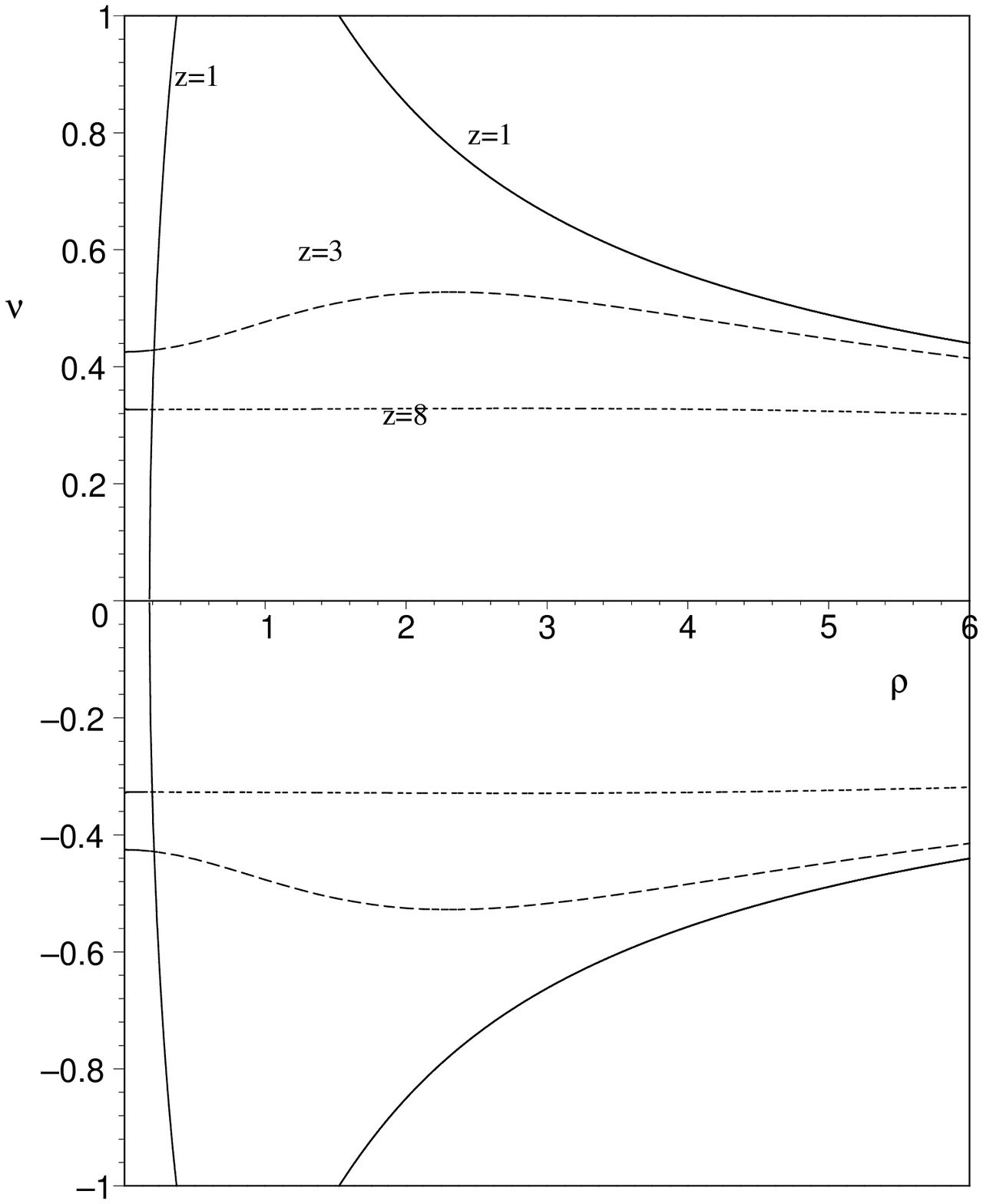}&\quad
\includegraphics[scale=0.35]{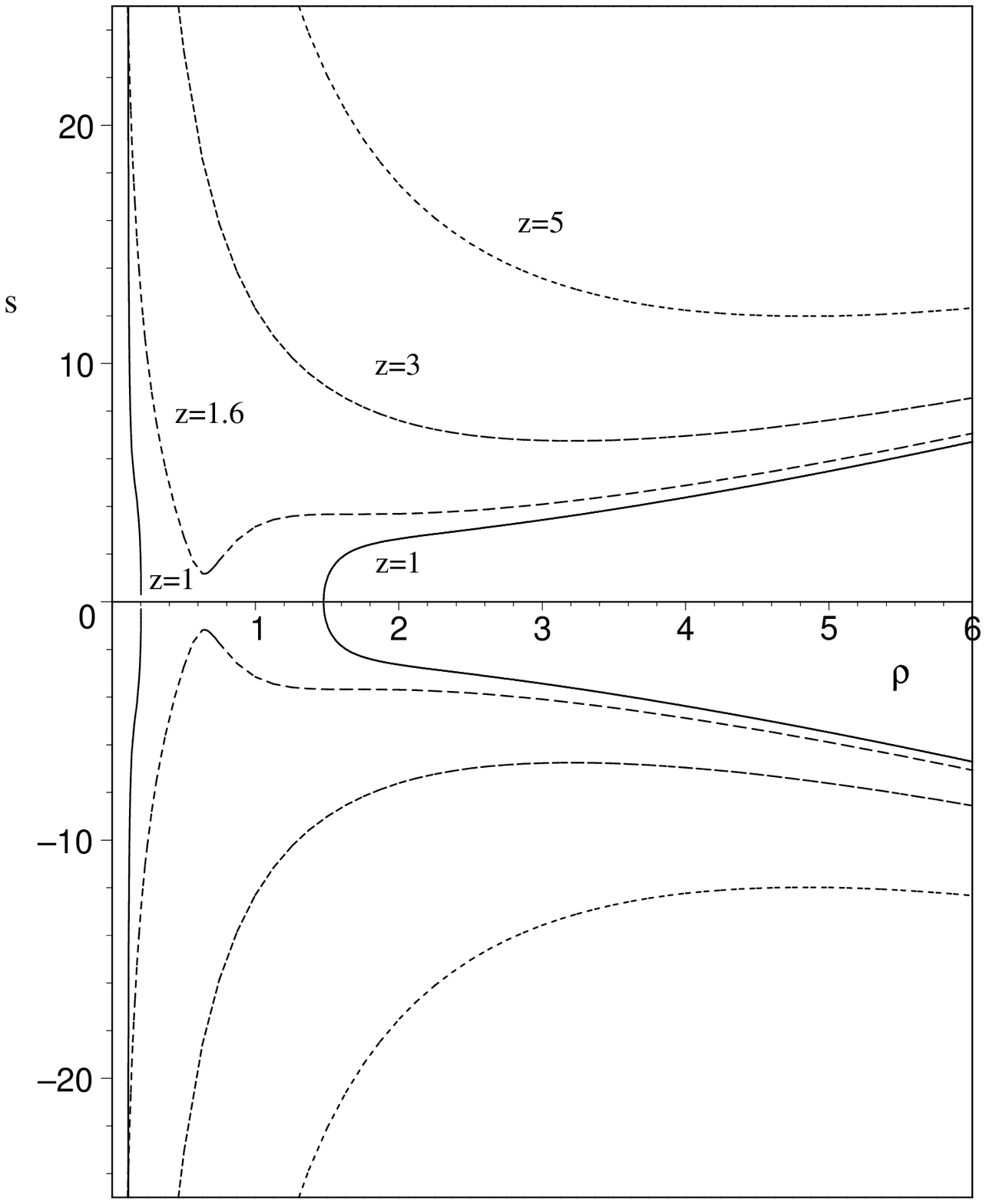}\\[.4cm]
\quad\mbox{(a)}\quad &\quad \mbox{(b)}\\[.6cm]
\includegraphics[scale=0.35]{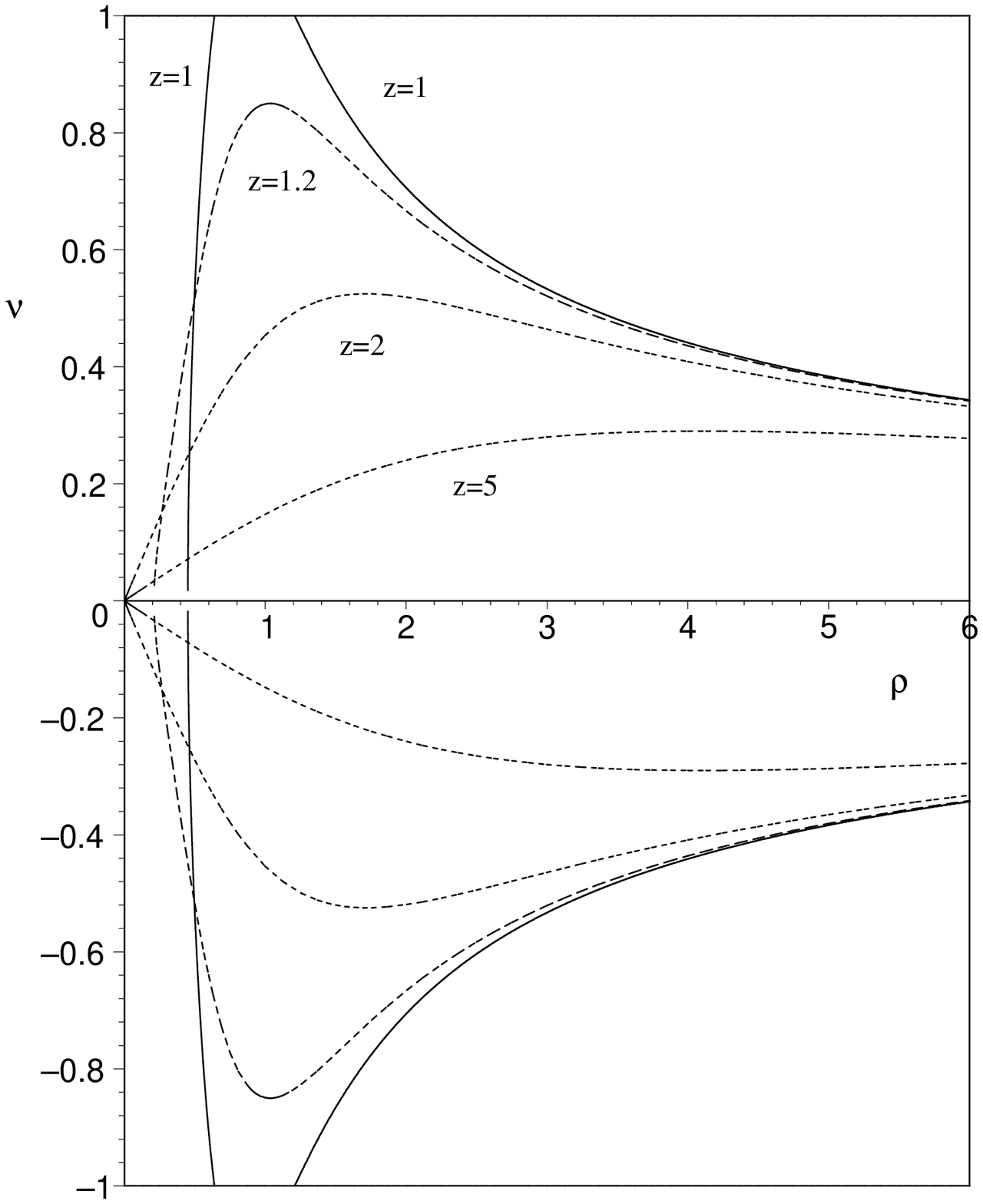}&\quad
\includegraphics[scale=0.35]{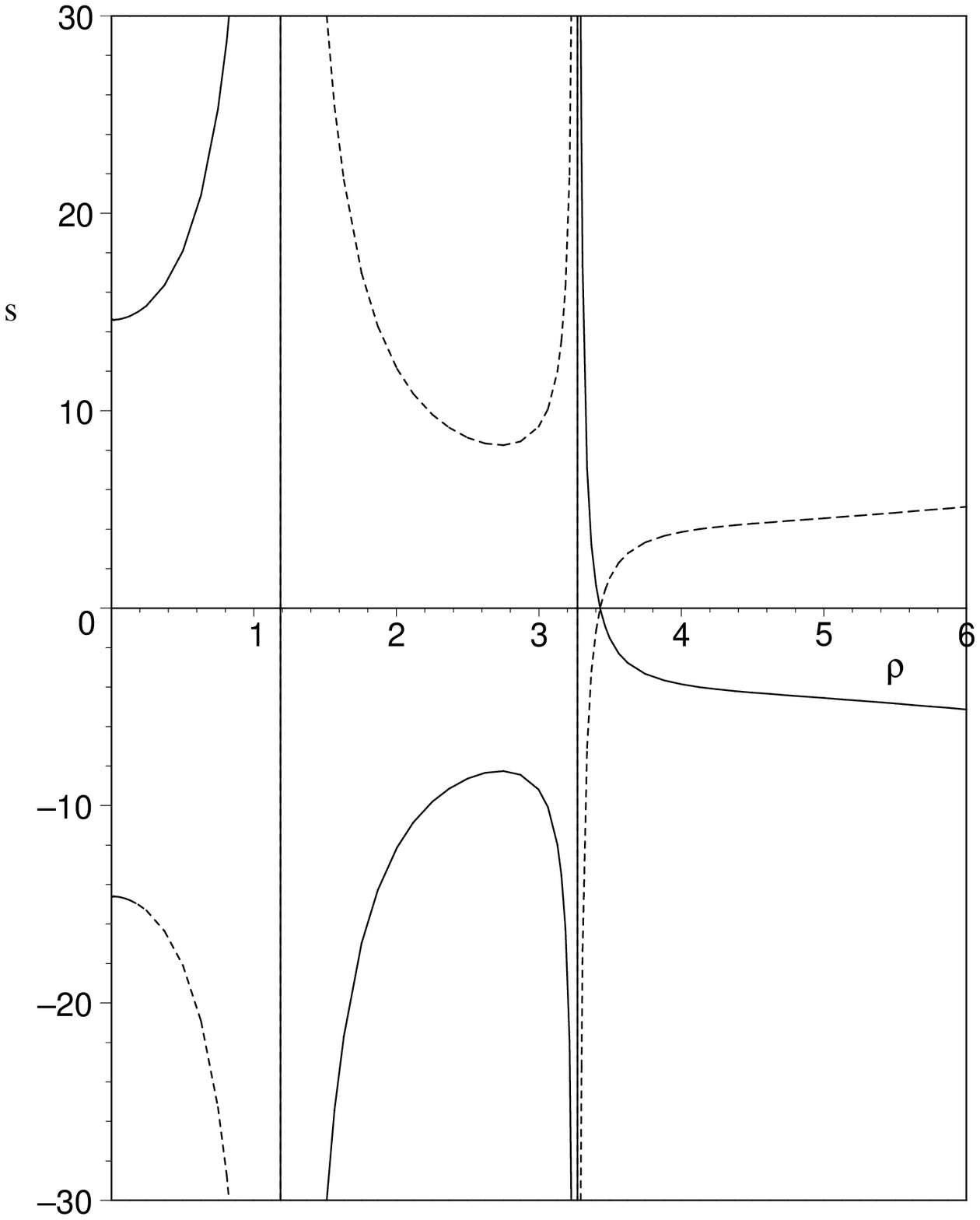}\\[.4cm]
\quad\mbox{(c)}\quad &\quad \mbox{(d)}
\end{array}
$\\
\end{center}
\caption{
In the case of the single Schwarzschild black hole, the linear velocity ${}^s\nu_{\pm}$ for co/counter-rotating circular orbits and the corresponding spin parameter ${\hat s}$ are evaluated on different planes $z=const$, and plotted in figures (a), (c) and figures (b), (d) respectively as functions of $\rho$ (and $M_{\rm S}=1$), 
for both CP (see figures (a), (b)) and P (see figures (c), (d)) supplementary conditions.
Solid, dotted and dashdotted lines correspond to $z=1,3,8$ respectively in figure (a); solid, dotted, dashed and dashdotted lines refer to the choices $z=1,1.6,3,5$ in figure (b) and $z=1,1.2,2,5$ in figure (c); 
in figure (d) we plot only the case $z=3$ as an example, with solid and dotted lines referring to co/counter-rotating orbits respectively.
}
\label{fig:5}
\end{figure}

\begin{figure} 
\typeout{*** EPS figure 6}
\begin{center}
$
\begin{array}{cc}
\includegraphics[scale=0.35]{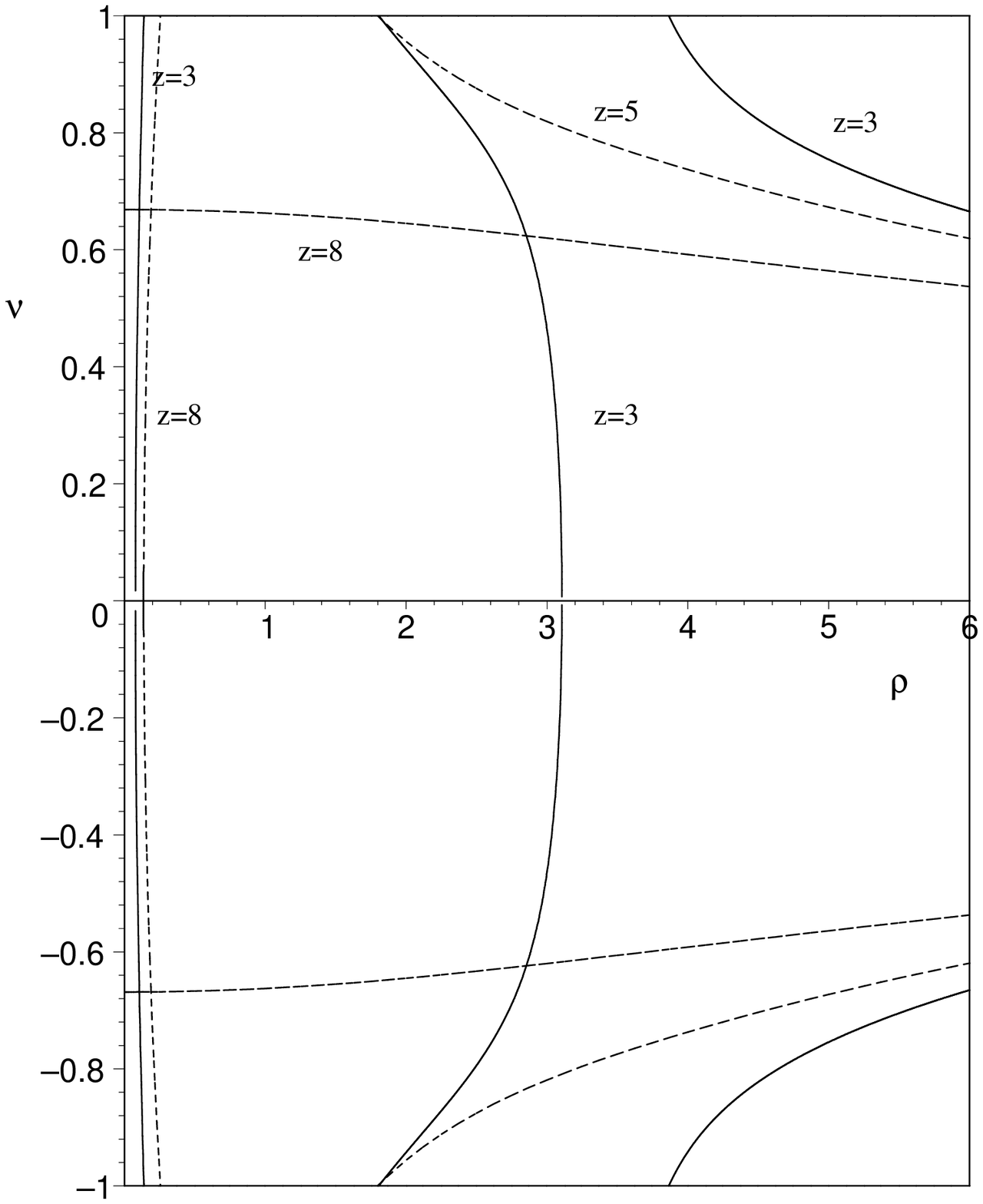}&\quad
\includegraphics[scale=0.35]{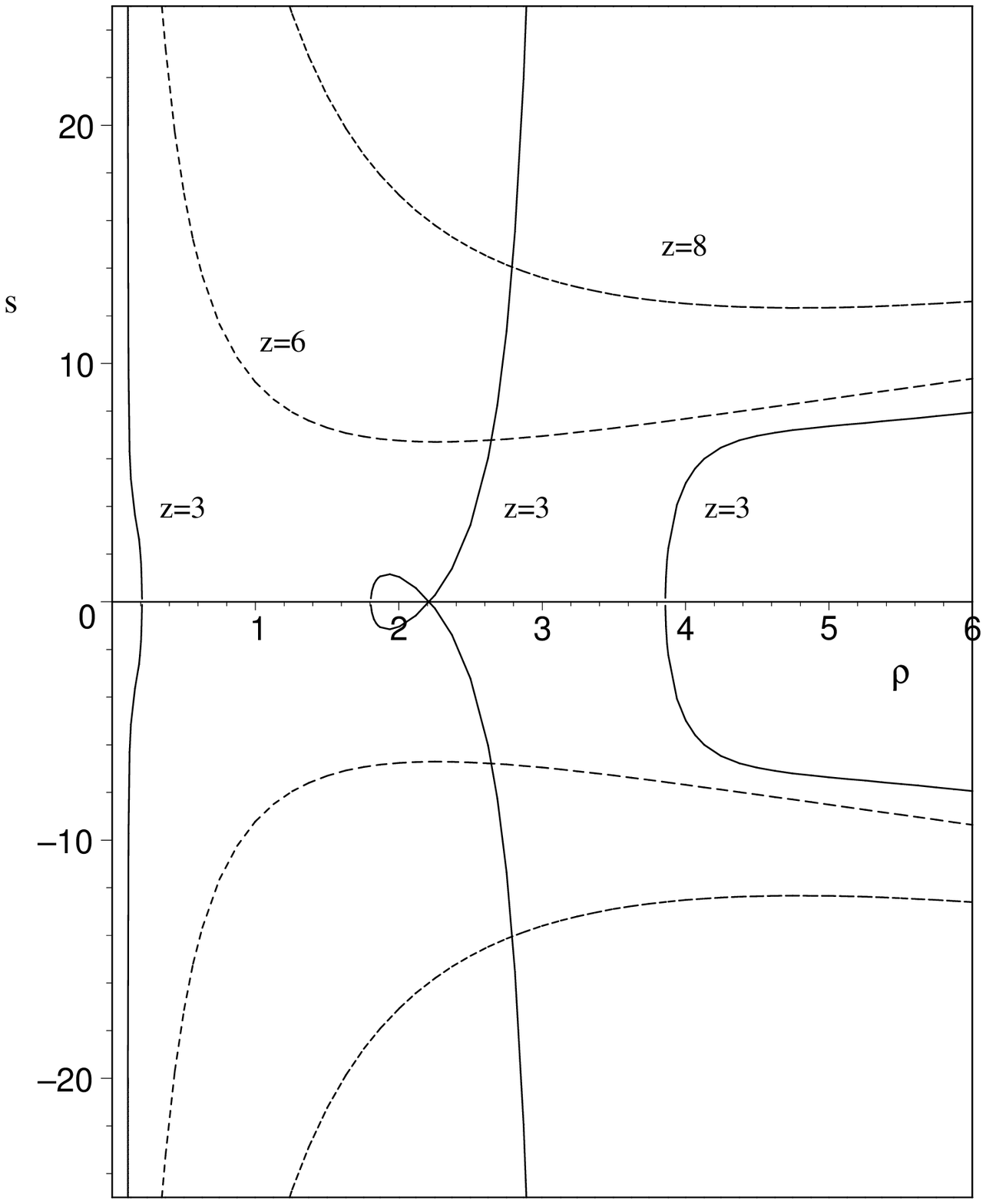}\\[.4cm]
\quad\mbox{(a)}\quad &\quad \mbox{(b)}\\[.6cm]
\includegraphics[scale=0.35]{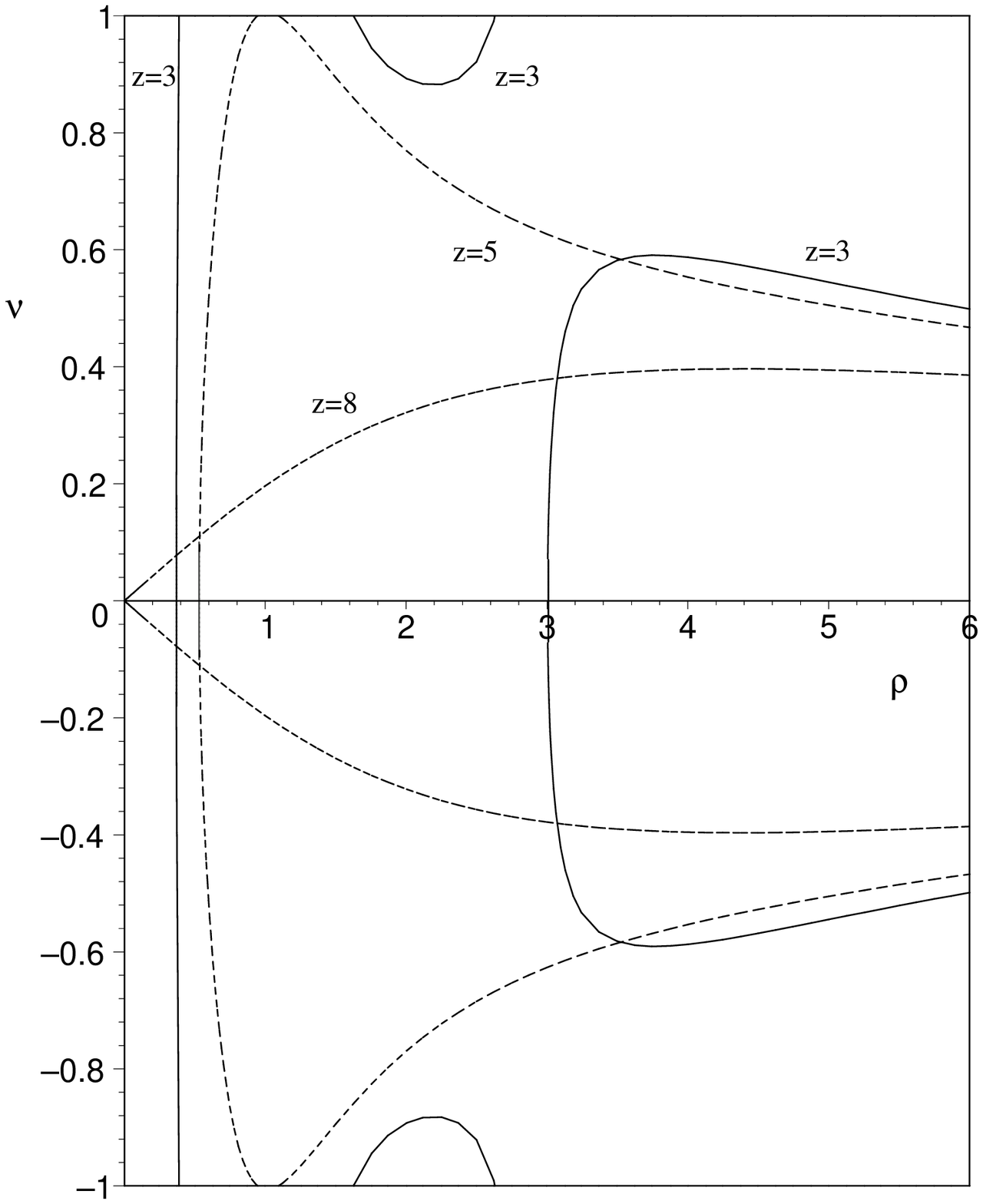}&\quad
\includegraphics[scale=0.35]{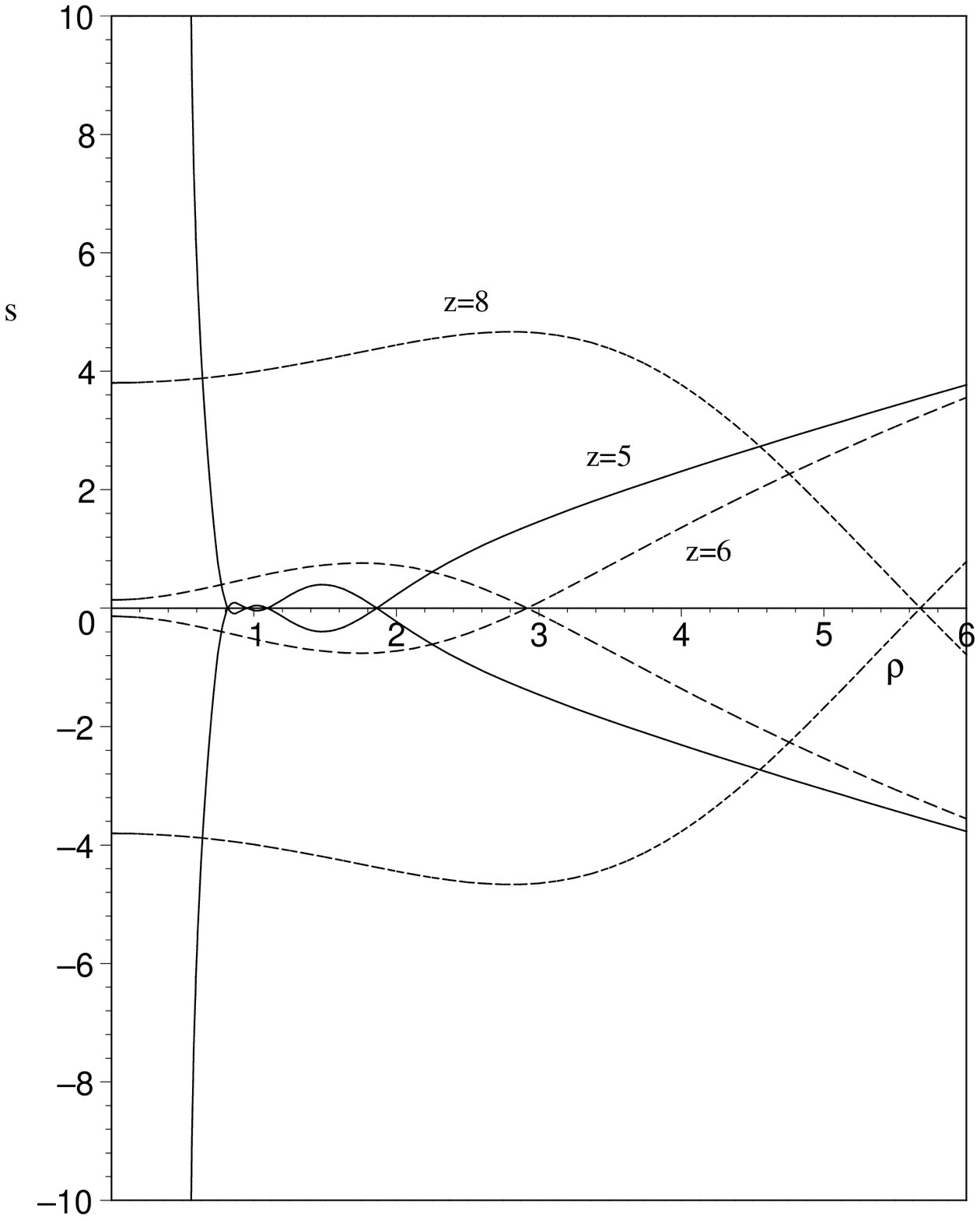}\\[.4cm]
\quad\mbox{(c)}\quad &\quad \mbox{(d)}
\end{array}
$\\
\end{center}
\caption{
In the case of two Schwarzschild black holes, the linear velocity ${}^s\nu_{\pm}$ for co/counter-rotating circular orbits and the corresponding spin parameter ${\hat s}$ are evaluated on different planes $z=const$, and plotted in figures (a), (c) and figures (b), (d) respectively as functions of $\rho$ (and $M_{\rm S}=1=m_{\rm S_b}$, $b=4$), 
for both CP (see figures (a), (b)) and P (see figures (c), (d)) supplementary conditions.
Solid, dotted and dashed lines correspond to $z=3,5,8$ respectively in figure (a), $z=3,6,8$ in figure (b), $z=3,5,8$ in figure (c) and $z=5,6,8$ in figure (d).
}
\label{fig:6}
\end{figure}

About the general case ($k_{\rm (lie)}{}_{\hat z}\not=0$), the behaviours of the linear velocities ${}^s\nu_{\pm}$ for co/counter-rotating circular orbits and of the corresponding spin parameter ${\hat s}$ are shown in Figures \ref{fig:3} to \ref{fig:6}, 
in the case of CP (Figures (a) and (b)) and P (Figures (c) and (d)) supplementary conditions, as functions of the radial coordinate $\rho$ and evaluated on different planes $z=const$.
In the case of T supplementary conditions, the relations defining the quantities ${}^s\nu_{\pm}$ and ${\hat s}$ are known only implicitly by means of equations~(\ref{ssolT}) and (\ref{ssolTfromnup}). 
The usefulness of the plots in Figures 3 to 6 is that of providing the values
of ${}^s\nu_\pm$ or that of the spin $s$ necessary to have a circular orbit
at a any given value of $\rho$ for fixed $z$.

\section{Conclusions}

Spinning test particles in  circular motion in static vacuum spacetimes belonging to the Weyl class have been discussed in detail in the framework of the 
Mathisson-Papapetrou approach supplemented by standard conditions.
In the limit of small spin and on particular symmetry hyperplanes, the orbit of the particle is close to a circular geodesic and
the difference in the angular velocities with respect to the geodesic value can be of arbitrary sign, 
corresponding to the two spin-up and spin-down orientations along the $z$-axis. 
For co-rotating and counter-rotating both spin-up (or both spin-down) test particles a nonzero gravitomagnetic \lq\lq clock effect'' 
appears under the same conditions. 
Applications to specific static Weyl spacetimes, corresponding to a single Chazy-Curzon particle and a Schwarzschild black hole as well as to two Chazy-Curzon particles and two Schwarzschild black holes, are discussed (mostly with the aid of plots) for the standard choices of supplementary conditions.

\newpage


\typeout{plots  $K_z\not= 0$}

\end{document}